\documentclass[11pt]{article}

\usepackage{amsmath}
\usepackage{amsfonts}
\usepackage{amssymb}
\usepackage{cite}
\usepackage{epsfig}
\usepackage{graphicx}
\usepackage[dvips]{color}

\newlength{\extraspace}
\newlength{\extraspaces}
\newcommand{\B}[1]{{\mathbb #1}}
\newcommand{\C}[1]{{\mathcal #1}}

\newcommand{\BF}[1]{{\mathbf #1}}

\newcommand{\beq}{\begin{equation}}
\newcommand{\eeq}{\end{equation}}
\newcommand{\beqn}{\begin{equation*}}
\newcommand{\eeqn}{\end{equation*}}
\newcommand{\bea}{\begin{eqnarray}}
\newcommand{\eea}{\end{eqnarray}}
\newcommand{\bean}{\begin{eqnarray*}}
\newcommand{\eean}{\end{eqnarray*}}
\newcommand{\nn}{\nonumber}

\newcommand{\Tr}{\mathop{\rm Tr}}

\newcommand{\half}{\frac 12}

\begin{document}

\thispagestyle{empty}

\begin{flushright}
{\sc HIP}-2009-05/TH\\
{arXiv:0903.3223[hep-th]}\\
March 2009\\
\end{flushright}
\vspace{.3cm}

\begin{center}
{\LARGE\bf{
BPS Vortices in Non-relativistic M2-brane
Chern-Simons-matter Theory
} }\\[15mm]

{\large 
Shinsuke Kawai${}^\dag$\footnote{e-mail: {\tt shinsuke.kawai(AT)helsinki.fi}}
and Shin Sasaki${}^{\dag\ddag}$\footnote{e-mail: {\tt shin.sasaki(AT)helsinki.fi}}}
\\[5mm]
${}^\dag$
{\it Helsinki Institute of Physics, University of Helsinki,
P.O.Box 64, Helsinki 00014 Finland}
\\[4mm]
${}^\ddag$
{\it Department of Physics, University of Helsinki, 
P.O.Box 64, Helsinki 00014 Finland}
\\[15mm]
{\bf Abstract}

\begin{center}
\begin{minipage}{14cm}
We study BPS vortices in the mass-deformed non-relativistic ${\C N}=6$ $U(N)_k\times U(N)_{-k}$ 
Chern-Simons-matter theory. 
We focus on the massive deformation that preserves the maximal ${\C N}=6$ supersymmetry, 
and consider a non-relativistic limit that carry 14 supercharges. 
In this non-relativistic field theory we find Jackiw-Pi type exact vortex solutions combined with $S^3$ fuzzy sphere geometry. We analyse their properties and show that they preserve 
one dynamical, one conformal and five kinematical supersymmetries among the full super
Schr\"odinger symmetry. 

\end{minipage}
\end{center}

\end{center}

\noindent

\vfill\noindent
PACS number(s): 11.10.Kk, 11.25.Yb, 11.27.+d

\newpage
\pagestyle{plain}
\setcounter{page}{1}
\tableofcontents

\section{Introduction}

Highly supersymmetric three dimensional conformal field theory has attracted much attention 
recently. 
A conformal theory having ${\C N}=8$ supersymmetry was constructed by
Bagger and Lambert \cite{Bagger:2006sk,Bagger:2007jr,Bagger:2007vi} and Gustavsson
\cite{Gustavsson:2007vu,Gustavsson:2008dy} and was proposed as a low-energy effective
theory describing the world-volume of two coincident M2 branes in M-theory. 
A salient feature of their construction is that it entails a so-called three-algebra.
There was a puzzle on how to generalize this model to include an arbitrary number of M2 branes;
this was elegantly solved by 
Aharony, Bergman, Jafferis and Maldacena \cite{Aharony:2008ug} (hereafter ABJM)
using a $U(N)\times U(N)$ Chern-Simons-matter theory at level $(k, -k)$, describing $N$
coincident M2 branes probing a transverse ${\B C}^4/{\B Z}_k$ orbifold space. 
The model has ${\C N}=6$ supersymmetry for generic $k$ but for $k=1$ and $2$ the 
supersymmetry is enhanced to ${\C N}=8$.  
The model is believed to have a gravity dual description which is M-theory on 
$AdS_4\times S^7/{\B Z}_k$.
In the 't Hooft limit of large $N$ and large $k$ with fixed $N/k$ this reduces to IIA string theory
on $AdS_4\times{\B C\B P}^3$.
This model was reformulated using the ${\C N}=2$ superspace formalism and further generalized 
in \cite{Benna:2008zy}.

Since the model of ABJM was proposed there has been a keen interest in constructing classical 
solutions in this model, such as BPS fuzzy-funnels \cite{Terashima:2008sy}, domain 
walls \cite{Hanaki:2008cu}, vortices and  Q-balls \cite{Arai:2008kv}, as well as time-dependent (non-BPS) fuzzy spheres \cite{Fujimori:2008ga}.
Solitonic solutions in the Bagger-Lambert-Gustavsson model have also been studied in 
\cite{Krishnan:2008zm,KIm:2008cp}; see also \cite{Jeon:2008zj,Jeon:2008bx,Bonelli:2008kh}.
These are particularly interesting from the M-theory viewpoint since they are expected to correspond to various configurations of membranes.

Apart from M-theory, three-dimensional Chern-Simons-matter theory appears in various models of
low-dimensional condensed matter systems (see \cite{Dunne:1998qy,Horvathy:2008hd} for reviews). 
While supersymmetry is not essential in this context, theories like ABJM are expected to provide various
examples of solvable toy models.
A new vogue in high energy theoretical physics is to apply the idea of AdS/CFT
duality, or gauge-gravity duality more generally, to unveil non-perturbative aspects of field theory models.
A practical approach for studying the physics of superconductivity \cite{Hartnoll:2008vx} and 
quantum Hall effect \cite{KeskiVakkuri:2008eb} in this context is to contemplate
an abelian Higgs model on an AdS blackhole geometry that reproduces desired boundary behaviour.
It is hoped that an ABJM-like set up can be used to construct D-brane configurations
that directly give rise to holographic descriptions of such physics 
\cite{Fujita:2009kw,Hikida:2009tp}.

In condensed matter field theory interesting physics usually arises in non-relativistic regime.
Recently, the non-relativistic version of the AdS/CFT correspondence
\cite{Son:2008ye,Balasubramanian:2008dm,Maldacena:2008wh,Adams:2008wt,Herzog:2008wg}
is actively investigated in a hope to open up possibilities to test the conjectured duality against
direct laboratory experiments. 
Motivated by this, as well as by the discrete light-cone quantisation of M-theory, non-relativistic 
limits of the ABJM model have been studied by several 
groups \cite{Nakayama:2009cz,Lee:2009mm}.
It has been found that different non-relativistic limits can be taken, with different
numbers of unbroken supersymmetries.

In this paper we study solitonic solutions in the non-relativistic version of the ABJM model.
We find vortex solutions, providing the first example of BPS solitonic solutions in this model.
It is known \cite{Arai:2008kv} that the relativistic mass-deformed ABJM model possesses 
Jackiw-Lee-Weinberg vortex solutions \cite{Jackiw:1990pr}.
While our analysis may be considered to be the non-relativistic counterpart, it is certainly not 
possible to take non-relativistic limits on the solution level as the structure of the supersymmetry
algebra and the shape of the potential change qualitatively in these limits.
We elaborate on various technicalities and construct exact solutions of abelian vortices, which
turn out to involve Jackiw-Pi solutions \cite{Jackiw:1990mb} as their subelement.
We then analyse the supersymmetric properties of these solutions and show that these are
exactly half-BPS with respect to the non-relativistic supersymmetry. 
As vortices are known to play key roles in the physics of superconductor and quantum Hall effect, 
we expect these solutions may serve as an exact toy example in the framework of AdS/CMP (condensed matter physics) correspondence.

The plan of this paper is as follows.
In the next section we collect known results of the relativistic ABJM model and its massive deformation.
In Section 3 we review non-relativistic limits of this theory, and in Section 4 we describe our construction of vortex solutions. We discuss supersymmetric properties of these solutions in Section 5, and conclude in Section 6 with discussions. 
In the Appendix we outline the derivation of the non-relativistic supersymmetry transformation rules that we use in Section 5.

\section{The ABJM model and its massive deformation}
\subsection{The massless model}

We start with the ABJM model \cite{Aharony:2008ug}, i.e. a Chern-Simons-matter theory of gauge group $U(N)\times U(N)$ at level $(k, -k)$, with matter fields belonging to the bi-fundamental representation of this group. 
The bosonic part of the action is 
\beq
S^{\rm bos}_{\rm ABJM}
=\int d^3x \Big({\C L}^{\rm bos}_{\rm kin}+{\C L}_{\rm CS}-V^{\rm bos}_{D}-V^{\rm bos}_{F}\Big),
\eeq
where
\bea
{\C L}^{\rm bos}_{\rm kin}&=&-\Tr\Big[
(D_\mu Z^{\hat A})^\dag(D^\mu Z^{\hat A})+(D_\mu W_{\check A})^\dag(D^\mu W_{\check A})
\Big],\\
{\C L}_{\rm CS}&=&\frac{k}{4\pi}\epsilon^{\mu\nu\lambda}\Tr\Big[
A_\mu\partial_\nu A_\lambda+\frac{2i}{3}A_\mu A_\nu A_\lambda
-\hat A_\mu\partial_\nu\hat A_\lambda-\frac{2i}{3}\hat A_\mu\hat A_\nu\hat A_\lambda\Big],
\label{eqn:CS}
\\
V^{\rm bos}_{D}
&=&\frac{4\pi^2}{k^2}\Tr\Big[
\left| Z^{\hat B} Z^\dag_{\hat B} Z^{\hat A}-Z^{\hat A} Z^\dag_{\hat B} Z^{\hat B}
-W^{\dag \check B}W_{\check B} Z^{\hat A}+Z^{\hat A} W_{\check B} W^{\dag \check B}\right|^2\nn\\
&&\;\;\;+\left| W^{\dag\check B}W_{\check B} W^{\dag\check A}
-W^{\dag\check A}W_{\check B} W^{\dag\check B}
-Z^{\hat B} Z^\dag_{\hat B} W^{\dag\check A}
+W^{\dag\check A}Z^\dag_{\hat B} Z^{\hat B}\right|^2\Big],
\eea
and 
\bea
V^{\rm bos}_{F}
&=&\frac{16\pi^2}{k^2}\Tr\Big[
\left|\epsilon_{\hat A\hat C}\epsilon^{\check B\check D}
W_{\check B}Z^{\hat C}W_{\check D}\right|^2
+\left|\epsilon^{\check A\check C}\epsilon_{\hat B\hat D}
Z^{\hat B} W_{\check C} Z^{\hat D}\right|^2\Big].
\eea
Here $A_\mu$, $\hat A_\mu$ are the $U(N)\times U(N)$ gauge fields, 
$Z^{\hat A}$, $W^{\dag\check A}$ ($\hat A=1, 2$, $\check A=3,4$) 
are complex scalar fields in the 
$U(N)\times U(N)$ bi-fundamental $({\BF N}, \bar{\BF N})$ representation,
the world-volume metric is $\eta_{\mu\nu}=(-1, +1, +1)$, and
$\epsilon$'s are completely antisymmetric and $\epsilon^{012}=1$, 
$\epsilon^{12}=1=-\epsilon_{12}$.
Our conventions closely follow those of \cite{Benna:2008zy} but we set the normalisation of 
the $U(N)$ generators to be $\Tr T^a T^b = \frac{1}{2} \delta^{ab}$.
The gauge covariant derivative is 
\begin{eqnarray}
& & D_{\mu} Z^{\hat A} = \partial_{\mu} Z^{\hat A} + i A_{\mu} Z^{\hat A} 
- i Z^{\hat A} \hat{A}_{\mu},
\end{eqnarray}
the gauge field strength is defined by 
\begin{eqnarray}
F_{\mu \nu} = \partial_{\mu} A_{\nu} - \partial_{\nu} A_{\mu}
+ i [A_{\mu}, A_{\nu}],
\end{eqnarray}
and similarly for $\hat{A}_{\mu}$.
The common $U(1)$ charge is fixed to $+1$.
The model exhibits a manifest $SU(2) \times SU(2) \times U(1)_R$ global symmetry. 
Under each $SU(2)$, $Z^{\hat A}$ and $W_{\check A}$ transform independently in the 
fundamental representation.
In addition to this manifest symmetry, there is an $SU(2)_R$ symmetry 
under which  $(Z^1, W^{\dagger 3})$ and $(Z^2, W^{\dagger 4})$ transform as doublets. 
It is argued in \cite{Aharony:2008ug} that the $SU(2) \times SU(2)$ global symmetry 
combined with the $SU(2)_R$ gives rise to an enhanced R-symmetry $SU(4)_R\simeq SO(6)_R$.
Hence for generic values of $k$ the model is endowed with $\mathcal{N} = 6$ 
supersymmetry (SUSY).
For $k=1$ and $2$ the SUSY is further enhanced to ${\C N}=8$.

We consider a trivial embedding of the world-volume in the space-time, 
namely, the world-volume coordinates $(x^0, x^1, x^2)$ are identified 
with the space-time coordinates $(X^0, X^1, X^2)$.
The four complex scalars $Z^{\hat A}, W^{\dagger\check A}$ represent the transverse 
displacement of the M2-branes along the eight directions 
$X^I \ (I = 3, \cdots, 10)$. 
The model is expected to describe $N$ coincident M2-branes probing 
${\B C}^4/{\B Z}_k$ in eleven dimensions, 
with the orbifolding symmetry ${\B Z}_k$ acting as 
$(Z^{\hat A}, W^{\dagger\check A}) \to e^{\frac{2\pi i}{k}} (Z^{\hat A}, W^{\dagger\check A})$.

Combining with the fermionic part, the massless ABJM model Lagrangian can be written in the 
$SU(4)$ invariant form as \cite{Benna:2008zy}
\begin{eqnarray}
\mathcal{L}_{\rm ABJM} &=& \mathcal{L}_{\rm CS} + \mathcal{L}_{\rm kin} + \mathcal{L}_{\rm Yuk} 
+ \mathcal{L}_{\rm pot},
\end{eqnarray}
where ${\C L}_{\rm CS}$ is (\ref{eqn:CS}) and 
\begin{eqnarray}
\mathcal{L}_{\rm kin} &=& 
-\Tr\left[
D_{\mu} Y^{\dagger}_A D^{\mu} Y^A + i \Psi^{\dagger A} \gamma^{\mu} 
D_{\mu} \Psi_A
\right], \\
\mathcal{L}_{\rm Yuk} &=& - \frac{2\pi i}{k} 
\mathrm{Tr} \left[
Y^{\dagger}_A Y^A \Psi^{\dagger B} \Psi_B - Y^A Y^{\dagger}_A \Psi_B 
\Psi^{\dagger B} - 2 Y^{\dagger}_A Y^B \Psi^{\dagger A} \Psi_B
\right. \nonumber \\
& & \qquad \qquad \left.
- \epsilon^{ABCD} Y^{\dagger}_A \Psi_B Y^{\dagger}_C \Psi_D 
+ \epsilon_{ABCD} Y^A \Psi^{\dagger B} Y^C \Psi^{\dagger D}
\right], \\
\mathcal{L}_{\rm pot} &=& \frac{4\pi^2}{3k^2} \mathrm{Tr} 
\left[
(Y^A Y^{\dagger}_A)^3 + (Y^{\dagger}_A Y^A)^3 + 4 Y^A Y^{\dagger}_B Y^C 
Y^{\dagger}_A Y^B Y^{\dagger}_C - 6 Y^A Y^{\dagger}_B Y^B Y^{\dagger}_A 
Y^C Y^{\dagger}_C
\right].
\end{eqnarray}
We have combined the two $SU(2)$ indices $\hat{A} = 1,2, \ \check{A} = 3,4$ into 
one $SU(4)$ index $A = 1, \cdots, 4$ and rewritten the fields
\begin{eqnarray}
& & Y^A = (Z^{\hat{A}}, W^{\dagger \check{A}} ), \quad 
Y^{\dagger}_A = ( Z^{\dagger}_{\hat{A}}, W_{\check{A}} ), \\
& & \Psi_A = (\Psi_{\hat{A}}, \Psi_{\check{A}} ), \quad 
\Psi^{\dagger A} = 
(
\Psi^{\dagger \hat{A}}, \Psi^{\dagger \check{A}}
).
\end{eqnarray}
The potential part can be written as a complete square form \cite{Lee:2009mm}
\begin{eqnarray}
\mathcal{L}_{\rm pot} &=& - V_{\rm pot} = - \frac{2}{3} \mathrm{Tr} 
\left[
W_A {}^{BC} W^{\dagger A} {}_{BC}
\right],
\end{eqnarray}
where 
\begin{eqnarray}
W_A {}^{BC} &=& G_A {}^{BC} - G_A {}^{CB}, 
\label{eqn:W}\\
G_A {}^{BC} &\equiv& - \frac{\pi}{k} 
\left\{
2 Y^B Y^{\dagger}_A Y^C + \delta_A {}^B (Y^C Y^{\dagger}_D Y^D - Y^D 
Y^{\dagger}_D Y^C)
\right\}.
\end{eqnarray}

The massless $\mathcal{N} = 6$ SUSY transformations are generated by six $(1+2)$-dimensional Majorana spinors $\epsilon_i$, $i=1,2,\cdots,6$.
We shall also use SUSY parameters $\omega_{AB}$ and $\omega^{AB}$ related to $\epsilon_i$
by
\beq
\omega_{AB}=\epsilon_i[\Gamma^i]_{AB},\qquad
\omega^{AB}=(\epsilon^i)[(\Gamma^i)^*]^{AB},
\eeq
where the $4\times 4$ matrices $\Gamma$ are chirally decomposed 6-dimensional 
$\Gamma$-matrices which can be written using the Pauli matrices as
\beq
\begin{array}{lll}
\Gamma^1=\sigma_2\otimes{\B I}_2, &
\Gamma^2=-i\sigma_2\otimes\sigma_3,& 
\Gamma^3=i\sigma_2\otimes\sigma_1, \\
\Gamma^4=-\sigma_1\otimes\sigma_2,&
\Gamma^5=\sigma_3\otimes\sigma_2, &
\Gamma^6=-i{\B I}_2\otimes\sigma_2.
\end{array}
\eeq
It is easy to see that
\beq
(\omega_{AB})^*=\omega^{AB},\qquad \omega^{AB}=\half\epsilon^{ABCD}\omega_{CD}.
\eeq
The $\mathcal{N} = 6$ SUSY transformations are then \cite{Terashima:2008sy}
\begin{eqnarray}
\delta Y^A &=& i \omega^{AB} \Psi_B, \\
\delta Y^{\dagger}_A &=& i \Psi^{\dagger B} \omega_{AB}, \\
\delta \Psi_A 
&=& - \gamma^{\mu} \omega_{AB} D_{\mu} Y^B 
- \omega_{BC} W_A {}^{BC} \label{SUSY_fermion}, \\
\delta \Psi^{\dagger A} 
&=& 
D_{\mu} Y^{\dagger}_B \gamma^{\mu} \omega^{AB} 
-\omega^{BC} W^{\dagger A} {}_{BC}, \\
\delta A_{\mu} &=& - \frac{2\pi}{k} 
\left(
Y^A \Psi^{\dagger B} \gamma_{\mu} \omega_{AB} + \omega^{AB} \gamma_{\mu} 
\Psi_A Y^{\dagger}_B
\right), \\
\delta \hat{A}_{\mu} &=& \frac{2\pi}{k} 
\left(
\Psi^{\dagger A} Y^B \gamma_{\mu} \omega_{AB} + \omega^{AB} \gamma_{\mu} 
Y^{\dagger}_A \Psi_B
\right).
\end{eqnarray}

\subsection{Massive deformation}

For constructing solitonic solutions one needs to introduce a mass scale into the action,
which is accomplished by massive deformation of the potential.
In this paper we follow the prescription of \cite{Hosomichi:2008jd,Gomis:2008vc} 
that preserves the maximal ${\C N}=6$ supersymmetry.

The $\mathcal{N} = 6$ massive deformation is obtained by modifying the ``superpotential"
$W_A {}^{BC}$ into $W_A {}^{BC} + \delta W_A {}^{BC}$, where
\beq
\delta W_A {}^{BC} = \frac{1}{2} 
(M_A {}^B Y^C - M_A {}^C Y^B), \qquad 
M_A {}^B = m\ \mathrm{diag} (1,1,-1,-1).
\eeq
Here, $m$ is a real parameter having the dimension of mass.
Note that $M^A{}_B=(M_A{}^B)^\dag=M_A{}^B$. 
Under the deformation the potential part is transformed into
\begin{eqnarray}
\mathcal{L}_{\rm pot}
&\rightarrow& - \frac{2}{3} \mathrm{Tr} 
\left[
(W_A {}^{BC} + \delta W_A {}^{BC}) (W^{\dagger A} {}_{BC} + \delta 
W^{\dagger A} {}_{BC})
\right] .
\end{eqnarray}
In components, the change of the Lagrangian due to the massive deformation is
\bea
&&\!\!\!\!\!\!\!\!\!\!\!
\delta \mathcal{L} =\mathrm{Tr} 
\left[
- m^2 Z^{\dagger}_{\hat A} Z^{\hat A} - m^2 W^{\dagger\check A} W_{\check A} 
+ \frac{4\pi m}{k} 
\left(
(Z^{\hat A} Z^{\dagger}_{\hat A})^2 - (W^{\dagger\check A} W_{\check A} )^2 
- (Z^{\dagger}_{\hat A} Z^{\hat A})^2 + (W_{\check A} W^{\dagger\check A})^2
\right)
\right].
\nonumber \\
\eea
This massive deformation breaks the $SU(4)_R$ symmetry down to 
$SU(2) \times SU(2) \times U(1) \times {\B Z}_2$.
The vacuum structure of this mass-deformed ABJM model is discussed in 
\cite{Gomis:2008vc}, where not only symmetric but also asymmetric phases are found.
The mass-deformed SUSY transformation law is obtained by replacing $W_A 
{}^{BC}$ with $W_A {}^{BC} + \delta W_A {}^{BC}$ in the prescription described 
at the end of the last subsection.

\section{Non-relativistic limit of the mass-deformed ABJM model}

The non-relativistic limit of the ABJM model was recently considered in 
\cite{Nakayama:2009cz,Lee:2009mm}.
Since this is essential for our discussion we shall review it here in detail.

For this purpose it is instructive to recover the speed of light $c$ and the 
Planck constant $\hbar$ in the Lagrangian\footnote{
The dimensions of constants and fields appearing in this section
in terms of mass $M$, length $L$ and time $T$ are:
$[\hbar] = M L^2 T^{-1}$, 
$[m] = M$, 
$[c] = L T^{-1}$, 
$[k] = L^{-1} T$,
$[Z^{\hat{A}}] = [W^{\dagger \check{A}}] = M^{1/2} L^{1/2} T^{-1/2}$, 
$[\psi_A] = M^{1/2} T^{-1/2}$,
$[A_{\mu}] = [\hat{A}_{\mu}] = L^{-1}$, 
$[A_t] = T^{-1}$, 
$[\omega] = L^{1/2}$.
}:
\begin{eqnarray}
\mathcal{L}_{\rm kin} &=& 
\mathrm{Tr}
\left[
\frac{1}{c^2}
D_t Y^A D_t Y^{\dagger}_A - D_i Y^A D_i Y^{\dagger}_A - \frac{m^2 c^2}{\hbar^2} 
Y^A Y^{\dagger}_A \right. 
\nonumber \\
& &
\left.
\qquad \qquad 
- i \Psi^{\dagger A} \gamma^{\mu} D_{\mu} \Psi_A + \frac{imc}{\hbar} 
\Psi^{\dagger \hat{A}} \Psi_{\hat{A}} - \frac{imc}{\hbar} 
\Psi^{\dagger \check{A}} \Psi_{\check{A}}
\right], \\
\mathcal{L}_{\mathrm{CS}} &=& \frac{k \hbar c}{4\pi} 
\epsilon^{\mu \nu \rho} 
\mathrm{Tr} 
\left[
A_{\mu} \partial_{\nu} A_{\rho} + \frac{2i}{3} A_{\mu} A_{\nu} A_{\rho}
- \hat{A}_{\mu} \partial_{\nu} \hat{A}_{\rho} - \frac{2i}{3} \hat{A}_{\mu} \hat{A}_{\nu} \hat{A}_{\rho}
\right], \\
\mathcal{L}_{\rm Yuk} &=& \frac{2 \pi i}{k \hbar c} 
\mathrm{Tr} \left[
Y^{\dagger}_A Y^A \Psi^{\dagger B} \Psi_B - Y^A Y^{\dagger}_A \Psi_B 
\Psi^{\dagger B} - 2 Y^{\dagger}_A Y^B \Psi^{\dagger A} \Psi_B
\right. \nonumber \\
& & \qquad \qquad \left.
- \epsilon^{ABCD} Y^{\dagger}_A \Psi_B Y^{\dagger}_C \Psi_D 
+ \epsilon_{ABCD} Y^A \Psi^{\dagger B} Y^C \Psi^{\dagger D}
\right], \\
\mathcal{L}_{D} &=&-V^{\rm bos}_D= - \mathrm{Tr}
\left[
\left|
\frac{2 \pi}{k \hbar c}
\left(
Z^{\hat B} Z^{\dagger}_{\hat B} Z^{\hat A} 
- Z^{\hat A} Z^{\dagger}_{\hat B} Z^{\hat B} 
- W^{\dagger\check B} W_{\check B} Z^{\hat A} 
+ Z^{\hat A} W_{\check B} W^{\dagger\check B} 
\right)
\right|^2 
\right.
\nonumber \\
& & \left.
+ \left|
\frac{2\pi}{k \hbar c} 
\left(
W^{\dagger\check B} W_{\check B} W^{\dagger\check A} 
- W^{\dagger\check A} W_{\hat B} W^{\dagger\check B} 
- Z^{\hat B} Z^{\dagger}_{\hat B} W^{\dagger\check A} 
+ W^{\dagger\check A} Z^{\dagger}_{\hat B} Z^{\hat B}
\right)
\right|^2
\right], \\
\mathcal{L}_{F} &=&-V^{\rm bos}_F
= - \frac{16 \pi^2}{k^2 \hbar^2 c^2} 
\mathrm{Tr} 
\left[
\left|
\epsilon_{\hat A\hat C} \epsilon^{\check B\check D} W_{\check B} Z^{\hat C} W_{\check D}
\right|^2
+ 
\left|
\epsilon^{\check A\check C} \epsilon_{\hat B\hat D} Z^{\hat B} W_{\check C} Z^{\hat D}
\right|^2
\right].
\end{eqnarray}
The mass contributions to the potential term are (note that the canonical mass terms have been
included in ${\C L}_{\rm kin}$)
\begin{eqnarray}
\mathcal{L}_m =  \frac{4 \pi m}{k \hbar^2} 
\mathrm{Tr} 
\left[
(Z^{\hat A} Z^{\dagger}_{\hat A})^2 - (Z^{\dagger}_{\hat A} Z^{\hat A})^2 
- (W^{\dagger\check A} W_{\check A})^2 + (W_{\check A} W^{\dagger\check A})^2
\right].
\end{eqnarray}
For the time component of the gauge potential we introduce
$A_0 \equiv \frac{1}{c} A_t$, $\hat A_0 \equiv \frac{1}{c}\hat A_t$.
The covariant derivative then becomes  
\begin{eqnarray}
D_i Z^{\hat A} &=& \partial_{i} Z^{\hat A} + i A_i Z^{\hat A} - i Z^{\hat A} \hat{A}_i, \\
D_t Z^{\hat A} &=& \partial_{t} Z^{\hat A} + i A_t Z^{\hat A} - i Z^{\hat A} \hat{A}_t. 
\end{eqnarray}

We focus on the symmetric sector of the vacua and decompose the (relativistic) scalar 
fields into the particle and antiparticle parts,
\begin{eqnarray}
Z^{\hat A} &=& \frac{\hbar}{\sqrt{2m}} 
\left(
e^{- i \frac{m c^2 t}{\hbar}} z^{\hat A} + e^{ i \frac{m c^2 t}{\hbar}} 
\hat{z}^{* \hat A}
\right), \\
Z^{\dagger}_{\hat A} &=& \frac{\hbar}{\sqrt{2m}} 
\left(
e^{i \frac{m c^2 t}{\hbar}} z^{\dagger}_{\hat A} + e^{- i \frac{m c^2 t}{\hbar}} 
\hat{z}^{* \dagger}_{\hat A}
\right), \\
W^{\dagger \check A} &=& \frac{\hbar}{\sqrt{2m}} 
\left(
e^{- i \frac{m c^2 t}{\hbar}} w^{\dagger\check A} + e^{ i \frac{m c^2 t}{\hbar}} 
\hat{w}^{* \dagger\check A}
\right), \\
W_{\check A} &=& \frac{\hbar}{\sqrt{2m}} 
\left(
e^{i \frac{m c^2 t}{\hbar}} w_{\check A} + e^{- i \frac{m c^2 t}{\hbar}} 
\hat{w}^{*}_{\check A}
\right).
\end{eqnarray}
Here, $z^{\hat A}$, $\hat z^{*\hat A}$ etc. are regarded as {\em non-relativistic} scalar fields. 
Let us keep the particle degrees of freedom $(z^{\hat A}, w^{\dagger\check A})$ and 
drop the antiparticle sector. 
Taking the non-relativistic limit amounts to sending $c, m\rightarrow\infty$ and considering the
leading orders.
The Chern-Simons term is not affected in this non-relativistic limit.
The kinetic part of the bosonic sector becomes
\begin{eqnarray}
\mathcal{L}^{\rm bos}_{\rm kin} &=& 
\mathrm{Tr} 
\left[
\frac{i \hbar}{2} 
\left(- z^{\dagger}_{\hat A} D_t z^{\hat A} + 
D_t z^{\hat A} \cdot z^{\dagger}_{\hat A}
\right)
+
\frac{\hbar^2}{2mc^2} 
D_t z^{\hat A} D_t z^{\dagger}_{\hat A} - \frac{\hbar^2}{2m} D_i z^{\hat A} D_i z^{\dagger}_{\hat A} 
\right. \nonumber \\
& & \left. \qquad 
+ \frac{i \hbar}{2} 
\left(
- w_{\check A} D_t w^{\dagger\check A} 
+ D_t w^{\dagger\check A} \cdot w_{\check A}
\right)
+ \frac{\hbar^2}{2mc^2} 
D_t w^{\dagger\check A} D_t w_{\check A} - \frac{\hbar^2}{2m} D_i w^{\dagger\check A} D_i w_{\check A} 
\right].
\end{eqnarray}
The terms $\frac{\hbar^2}{2mc^2} |D_t z^{\hat A}|^2$, 
$\frac{\hbar^2}{2mc^2} |D_t w^{\dag\check A}|^2$ are sub-leading in the limit 
$c,m \to \infty$. 
The potential terms $\mathcal{L}_D$ and $\mathcal{L}_F$ are also of sub-leading order. 
Nontrivial contributions in the potential come from the mass dependent part
\begin{eqnarray}
\mathcal{L}_m =  \frac{\pi \hbar^2}{k m} 
\mathrm{Tr} 
\left[
(z^{\hat A} z^{\dagger}_{\hat A})^2 - (z^{\dagger}_{\hat A }z^{\hat A})^2 
- (w^{\dagger\check A} w_{\check A})^2 + (w_{\check A} w^{\dagger\check A})^2
\right].
\end{eqnarray}
Assembling the terms up to $\mathcal{O}(1/c^2)$ we find (the bosonic part of) the Lagrangian
for the non-relativistic massive ABJM model in the symmetric phase:
\begin{eqnarray}
\mathcal{L}^{\rm NR, bos}_{\rm ABJM} &=& 
\frac{k\hbar c}{4\pi} 
\epsilon^{\mu \nu \lambda} \mathrm{Tr}
\left[
A_{\mu} \partial_{\nu} A_{\lambda} + \frac{2i}{3} A_{\mu} A_{\nu} 
A_{\lambda}
- \hat{A}_{\mu} \hat{A}_{\nu} \hat{A}_{\lambda} - \frac{2i}{3} 
\hat{A}_{\mu} \hat{A}_{\nu} \hat{A}_{\lambda}
\right]\nn\\
& & + \mathrm{Tr} 
\left[
\frac{i \hbar}{2} 
\left(- z^{\dagger}_{\hat A} D_t z^{\hat A} + D_t z^{\hat A} \cdot z^{\dagger}_{\hat A}\right)
- \frac{\hbar^2}{2m} D_i z^{\hat A} D_i z^{\dagger}_{\hat A} 
\right. \nonumber \\
& &  \qquad 
+ \frac{i \hbar}{2} 
\left(
- w_{\check A} D_t w^{\dagger\check A} 
+ D_t w^{\dagger\check A} \cdot w_{\check A}
\right)
- \frac{\hbar^2}{2m} D_i w^{\dagger\check A} D_i w_{\check A}  
\nonumber \\
& &  \qquad \left.
+
\frac{\pi \hbar^2}{k m} 
\left\{
(z^{\hat A} z^{\dagger}_{\check A})^2 - (z^{\dagger}_{\hat A} z^{\hat A})^2
- (w^{\dagger\check A} w_{\check A})^2 + (w_{\check A} w^{\dagger\check A})^2
\right\}
\right].
\label{eqn:NRABJMlag}
\end{eqnarray}

The equations of motion of the non-relativistic theory are read off from the Lagrangian.
For the scalar fields we find
\begin{eqnarray}
& & i \hbar D_t z^{\hat A} 
= - \frac{\hbar^2}{2m} D_i^2 z^{\hat A} 
-
\frac{2\pi \hbar^2}{k m} 
(z^{\hat B} z^{\dagger}_{\hat B} z^{\hat A} - z^{\hat A} z^{\dagger}_{\hat B} z^{\hat B}), \\
& & i \hbar D_t w^{\dagger\check A} = - \frac{\hbar^2}{2m} D_i^2 w^{\dagger\check A}
+
\frac{2\pi \hbar^2}{km} 
(w^{\dagger\check B} w_{\check B} w^{\dagger\check A} 
- w^{\dagger\check A} w_{\check B} w^{\dagger\check B}).
\end{eqnarray}
These are gauged non-linear Schr\"{o}dinger equations.
The gauge field equations of motion (the Gauss law constraints) are
\begin{eqnarray}
E_i &=&  \epsilon_{ij} J^j, 
\label{Gauss1}
\\
\frac{k \hbar c}{2\pi} B &=& \hbar c (z^{\hat A} z^{\dagger}_{\hat A} 
+ w^{\dagger\check A} w_{\check A}), 
\label{Gauss2}
\\
\hat{E}_i &=&  \epsilon_{ij} \hat{J}^j, 
\label{Gauss3}
\\
\frac{k \hbar c}{2\pi} \hat{B} 
&=& \hbar c (z^{\dagger}_{\hat A} z^{\hat A} + w_{\check A} w^{\dagger\check A}),
\label{Gauss4}
\end{eqnarray}
where
$\epsilon^{0ij} \equiv \epsilon^{ij}$,
$E_j \equiv F_{0j}$, $B \equiv F_{12}$,
$\hat{E}_j \equiv \hat{F}_{0j}$, $\hat{B} \equiv \hat{F}_{12}$
and
\bea
J^i &=& - \frac{i \hbar \pi}{k m c}
\left(
z^{\hat A} D_i z^{\dagger}_{\hat A} - D_i z^{\hat A} \cdot z^{\dagger}_{\hat A} 
+ w^{\dagger\check A} D_i w_{\check A} - D_i w^{\dagger\check A} \cdot w_{\check A}
\right), \\
\hat{J}^i &=& \frac{i \hbar \pi}{k m c}
\left(
z^{\dagger}_{\hat A} D_i z^{\hat A} - D_i z^{\dagger}_{\hat A} \cdot z^{\hat A} 
+ w_{\check A} D_i w^{\dagger\check A} - D_i w_{\check A} \cdot w^{\dagger\check A}
\right),
\eea
are the matter currents.
There is a $U(1)$ global symmetry
$(z^{\hat A}, w^{\dagger\check A}) \rightarrow e^{i \alpha} (z^{\hat A}, w^{\dagger\check A})$.
The corresponding Noether charge is
\beq
Q = - \int \! d^2 x \mathrm{Tr} 
\left[
z^{\dagger}_{\hat A} z^{\hat A} + w_{\check A} w^{\dagger\check A}
\right].
\eeq

Likewise, the non-relativistic limit of the fermionic part can be taken by decomposing
the fermions into the particle and antiparticle parts and then discarding (say) the
antiparticle part. 
We abide by the supersymmetry and shall keep the particle part of the spinor $\Psi^A$, 
which is \cite{Lee:2009mm}
\begin{eqnarray}
\Psi_A &=& \sqrt{\hbar c} 
(u_{+} \psi_{-A} (t, \vec{x}) + u_{-} \psi_{+A} (t, \vec{x})) e^{- i 
\frac{mc^2}{\hbar} t} \nonumber \\
&=& \sqrt{\frac{\hbar c}{2}} 
\left(
\begin{array}{c}
\psi_{-A} + \psi_{+A} \\
- i \psi_{-A} + i \psi_{+A}
\end{array}
\right)
e^{- i \frac{mc^2}{\hbar} t}.
\label{eqn:FermDcp}
\end{eqnarray}
The basis $u_\pm$ are mutually orthogonal two-component constant vectors
\beq
u_\pm\equiv
\frac{1}{\sqrt 2}\left(\begin{array}{c}1 \\ \mp i\end{array}\right),
\label{eqn:upm}
\eeq
and $\psi_{\pm A}$ are one-component spinors with dimension 
$[\psi] = L^{-3/2} T^{1/2}$.
The fermionic part of the kinetic term then becomes 
\begin{eqnarray}
\mathcal{L}^{\rm ferm}_{\rm kin} &=& 
\mathrm{Tr} 
\left[
\hbar c 
\bar{\psi}_{+}^A 
\left(
\frac{i}{c} D_t \psi_{-A} 
- i D_{-} \psi_{+A}
\right) 
+2mc^2\bar\psi_+^{\hat A}\psi_{-\hat A}
\right.
\nonumber \\
& & \qquad \left.
+ \hbar c \bar{\psi}_{-}^A 
\left(
\frac{i}{c} 
D_t \psi_{+A} 
- i D_{+} \psi_{-A}
\right)
+2mc^2\bar\psi_-^{\check A}\psi_{+\check A} 
\right].
\end{eqnarray}
The equations of motion up to $\mathcal{O} (c^0)$ are 
\begin{eqnarray}
& & i \hbar D_t \psi_{-A} + 2m c^2 \delta_{A}^{\hat A} \psi_{-\hat A} - i \hbar c D_{-} 
\psi_{+A} = 0, \label{fermioneq1}\\
& & i \hbar D_t \psi_{+A} + 2m c^2 \delta_{A}^{\check A}\psi_{+\check A} - i \hbar c D_{+} 
\psi_{-A} = 0.\label{fermioneq2}
\end{eqnarray}
Using these equations of motion half of the fermionic degrees of freedom can be dropped.

Finally, the Yukawa term becomes
\bea
{\C L}_{\rm Yuk}&=&
\frac{\pi\hbar^2}{km}\Tr\left[
y_A^\dag y^A(\bar\psi^B_+\psi_{-B}-\bar\psi^B_-\psi_{+B})
-\bar\psi^B_+y^Ay^\dag_A\psi_{-B}+\bar\psi^B_-y^Ay^\dag_A\psi_{+B}\right.\nn\\
&&\qquad 
+2\bar\psi^B_+y^Ay^\dag_B\psi_{-A}-2\bar\psi^B_-y^Ay^\dag_B\psi_{+A}
-2y^\dag_Ay^B(\bar\psi^A_+\psi_{-B}+\bar\psi^A_-\psi_{+B})\\
&&\qquad
\left.
+\epsilon^{ABCD}(y^\dag_A\psi_{-B}y^\dag_C\psi_{+D}-y^\dag_A\psi_{+B}y^\dag_C\psi_{-D})
-\epsilon_{ABCD}(y^A\bar\psi^B_+y^C\psi^D_--y^A\bar\psi^B_-y^C\psi^D_+)
\right],\nn
\eea
where we have denoted the particles collectively as
$y^A=(z^{\hat A}, w^{\dag\check A})$, 
$y^\dag_A=(z^\dag_{\hat A}, w_{\check A})$. 
The Yukawa term is subleading and does not contribute to the fermion equations of motion
(\ref{fermioneq1}), (\ref{fermioneq2}).

\section{The BPS equations and the vortex solutions}

Now let us find vortex solutions that saturate the BPS bound in this setup.
To find codimension two BPS solutions, we drop the fermion parts 
and consider static configurations.
The Hamiltonian of the system is the conserved Noether charge for the 
gauge covariant time-translation \cite{Leblanc:1992wu}
\begin{eqnarray}
& & \delta Z^{\hat A} = \epsilon D_0 z^{\hat A}, 
\quad \delta w^{\dagger\check A} = \epsilon D_0 w^{\dagger\check A}, \\
& & \delta A_0 = \delta \hat{A}_0 = 0, \quad \delta A_i = \epsilon E_i, 
\quad \delta \hat{A}_i = \epsilon \hat{E}_i.
\end{eqnarray}
The Hamiltonian density is given by 
\begin{eqnarray}
\mathcal{H} &=& \mathrm{Tr} 
\left[
\frac{\hbar^2}{2m} |D_i z^{\hat A}|^2 + \frac{\hbar^2}{2m} |D_i w^{\dagger\check A}|^2
\right. \nonumber \\
& & \qquad \left.
-
\frac{\pi \hbar^2}{km} 
\left\{
(z^{\hat A} z^{\dagger}_{\hat A})^2 - (z^{\dagger}_{\hat A} z^{\hat A})^2 
- (w^{\dagger\check A} w_{\check A})^2 + (w_{\check A} w^{\dagger\check A})^2
\right\}
\right].
\end{eqnarray}
In order to perform the Bogomol'nyi completion it is convenient to use the relation
\begin{eqnarray}
[D_i, D_j] z^{\hat A} = i (F_{ij} z^{\hat A} - z^{\hat A} \hat{F}_{ij}).
\end{eqnarray}
Using this relation and the Gauss law constraints, and writing 
$D_\pm\equiv D_1 \pm i D_2$,  we find 
that the energy functional simplifies to 
\begin{eqnarray}
E &=& \int \! d^2 x \ \mathcal{H} 
\nonumber \\
&=& \int \! d^2 x \ 
\mathrm{Tr} 
\left[
\frac{\hbar^2}{2m}
\left|
D_{-} z^{\hat A}
\right|^2
+ \frac{\hbar^2}{2m}
\left|
D_{+} w^{\dagger\check A}
\right|^2
\right]
+ \frac{\hbar^2}{2m} \int \! d^2 x \ S.
\end{eqnarray}
The second term is a surface term evaluated at the boundary 
\begin{eqnarray}
\int \! d^2 x \ S 
&=& - i \int \! d^2 x \ 
\left\{
\partial_1 \mathrm{Tr} \left[ z^{\hat A} D_2 z^{\dagger}_{\hat A}  \right]
- \partial_2 \mathrm{Tr} \left[ z^{\hat A} D_1 z^{\dagger}_{\hat A} \right]
- \partial_1 \mathrm{Tr} 
\left[
w^{\dagger\check A} D_2 w_{\check A}
\right]
+ \partial_2 
\mathrm{Tr} 
\left[
w^{\dagger\check A} D_1 w_{\check A}
\right]
\right\}
\nonumber \\
&=& - i \oint \! d x^i \ \mathrm{Tr} 
\left[
z^{\hat A} D_i z^{\dagger}_{\hat A} - w^{\dagger\check A} D_i w_{\check A}
\right].
\end{eqnarray}
Now, for a finite energy configuration the fields settle down to their vacua at infinity.
Then 
\begin{eqnarray}
\left. D_i z^{\hat A} \right|_{\mathrm{boundary}} 
=\left. D_i w_{\check A} \right|_{\mathrm{boundary}} = 0,
\end{eqnarray}
and the surface term vanishes.
We may conclude that the BPS bound is given by
\begin{eqnarray}
E = \int \! d^2 x \ 
\mathrm{Tr} 
\left[
\frac{\hbar^2}{2m}
\left|
D_{-} z^{\hat A}
\right|^2
+ \frac{\hbar^2}{2m} 
\left|
D_{+} w^{\dagger\check A}
\right|^2
\right] \ge 0,
\end{eqnarray}
which is saturated when both 
\begin{eqnarray}
D_{-} z^{\hat A} = 0, \quad D_{+} w^{\dagger\check A} = 0,
\label{BPSeq}
\end{eqnarray}
are satisfied.
These are the BPS vortex equations.

Let us find a solution to these equations.
The simplest solution is just a configuration that the scalars are proportional to 
the unit matrix $z^{\hat{A}}, w^{\dagger \check{A}} \propto \mathbf{1}_{N \times N}$ and $A_i = \hat{A}_i$.
In this case, the equations become trivial. 
The scalars and the gauge fields are determined by a (anti)holomorphic function of 
$z = x^1 + i x^2$. This configuration is possible even for the $N=1$ case.

Besides this trivial solution, we may find non-trivial, non-singular 
solutions specific to the multiple M2-brane configuration.
Although it is difficult to solve the matrix valued equation (\ref{BPSeq}) 
together with the gauge field equations (\ref{Gauss1})-(\ref{Gauss4}) in 
general, we may find solutions by assuming an ansatz that simplifies the equations:

\begin{eqnarray}
z^{\hat A} (x) = \psi_z (x) S^{I}, \quad w^{\dagger\check A} (x) = \psi_w(x) S^{I}, 
\quad A_i (x) = a_i (x) S^{I} S^{\dagger}_I, \quad 
\hat{A}_i (x) = a_i (x) S^{\dagger}_I S^I.
\label{ansatz}
\end{eqnarray}
Here $\psi_z (x)$, $\psi_w(x)$, and $a_i (x)$ are ordinary (not 
matrix-valued) functions
and $S^I$ are constant matrices.
In the first and second expressions the indices are understood to be
$\hat A=(1,2)\leftrightarrow I=(1,2)$, $\check A=(3,4)\leftrightarrow I=(1,2)$. 
The matrices $S^I \ (I=1,2)$ are the $N \times N$ ``vacuum matrices''
in the form \cite{Gomis:2008vc} 
\beq
(S^{\dagger}_1)_{mn} = \sqrt{m -1} \delta_{mn}, \quad 
(S^{\dagger}_2)_{mn} = \sqrt{N-m} \delta_{m+1,n}.
\eeq
It is easy to show that
\begin{eqnarray}
& & S^I = S^J S^{\dagger}_J S^I - S^I S^{\dagger}_J S^J, \\
& & S^{\dagger}_I = S^{\dagger}_I S^J S^{\dagger}_J 
- S^{\dagger}_J S^J S^{\dagger}_I, \\
& & \mathrm{Tr} S^I S^{\dagger}_I = \mathrm{Tr} S^{\dagger}_I S^I = N (N-1).
\end{eqnarray}
The BPS equations (\ref{BPSeq}) then reduce to 
\begin{eqnarray}
(\mathcal{D}_1-i \mathcal{D}_2) \psi_{z} (x) = 0,
\qquad (\mathcal{D}_1+i \mathcal{D}_2) \psi_{w} (x) = 0,
\end{eqnarray}
where
$\mathcal{D}_i \equiv \partial_i + i a_i$. 
These are in fact the vortex equations of Jackiw and Pi \cite{Jackiw:1990mb}.

Let us for simplicity set $w^{\dag\check A}=0$ 
and solve the equations for $z^{\hat A}$, $A_i$ and $\hat A_i$.
We call this solution ``BPS-I." 
Geometrically, this is a configuration of M2-branes polarized into a fuzzy $S^3$.
The physical radius of the fuzzy $S^3$ is 
evaluated as 
\begin{eqnarray}
R^2 = \frac{2}{N T_{M2}} \mathrm{Tr} 
\left[Z^{\hat A} Z^{\dagger}_{\hat A} \right]
= \frac{N-1}{T_{M2}} \frac{|\psi_z|^2}{m},
\end{eqnarray}
where $T_{M2}$ is the tension of an M2-brane.
Note that in the case of $N=1$, the fuzzy sphere collapses into zero size and there 
are no non-trivial solutions. 
Our solutions may be regarded as an embedding of the Jackiw-Pi abelian vortices in the
non-relativistic ABJM model (see also discussions in Section 6).
These solutions are specific to the multiple M2-branes. 
The size of the fuzzy sphere is related to the $U(1)$ charge of the vortices, 
as explained below.

It is well known that the Jackiw-Pi vortex equation allows exact solutions.
The Gauss law constraint for the ansatz (\ref{ansatz}) is 
\beq
b = f_{12},
\label{Gauss}
\eeq
where $b = \frac{2\pi}{k} |\psi_z|^2$ and 
$f_{ij} \equiv \partial_i a_j - \partial_j a_i$.
Changing the variables 
\begin{eqnarray}
\psi_z (x) = e^{i \theta (x)} \rho^{\frac{1}{2}} (x), \quad (\theta, \rho \in {\B R}),
\end{eqnarray}
the BPS equation becomes 
\begin{eqnarray}
(\mathcal{D}_1 - i \mathcal{D}_2) \psi &=& 
\left[
i \partial_1 \theta \rho^{1/2} + \frac{1}{2} \rho^{-1/2} \partial_1 \rho 
+ i a_1 \rho^{1/2} 
\right. \nonumber \\
& & \qquad \qquad \left.
+ \partial_2 \theta \rho^{1/2} - \frac{i}{2} \rho^{-1/2} \partial_2 \rho 
+ a_2 \rho^{1/2}
\right] e^{i \theta} = 0,
\end{eqnarray}
giving a pair of equations
\begin{eqnarray}
a_i (x) = - \partial_i \theta + \frac{1}{2} \epsilon_{ij} \partial^j \ln \rho.
\label{BPS_sol1}
\end{eqnarray}
Substituting these into the Gauss law constraint, we have the Liouville equation
\begin{eqnarray}
\nabla^2 \ln \rho = -\frac{4 \pi }{k} \rho,
\end{eqnarray}
which may be solved by
\begin{eqnarray}
\rho (x) = \frac{k}{2\pi } \nabla^2 \ln 
\left(
1 + |f(z)|^2
\right),
\label{BPS_sol2}
\end{eqnarray}
where $f(z)$ is a holomorphic function of $z = x_1 + i x_2$.
The $U(1)$ Noether charge for this configuration is
\begin{eqnarray}
Q &=& -  N (N-1) \int \! d^2 x \rho 
\nonumber \\
&=& -N (N-1) \frac{k}{2\pi} (2\pi) \int \! dr \
\left(
r \frac{\partial^2}{\partial r^2} + \frac{\partial}{\partial r} 
\right) 
\ln (1 + |f|^2),
\end{eqnarray}
where $r = |z|$. 
This is proportional to the magnetic charge via the relation 
(\ref{Gauss}). As is well known, this fact is specific for Chern-Simons vortices.

Particularly simple examples of vortex profiles are obtained by 
choosing the holomorphic function to be
\begin{eqnarray}
f (z) = \left(
\frac{z_0}{z}
\right)^n, \quad n \in{\B Z},
\label{eqn:PowerProfile}
\end{eqnarray}
where $z_0$ is a complex constant.
In this case
\begin{eqnarray}
\rho (x) = \frac{k}{2 \pi} \frac{4 n^2}{r_0^2}
\frac{\left(\frac{r}{r_0}\right)^{2 (n-1)}}
{\left[
1 + \left(\frac{r}{r_0}\right)^{2n}
\right]^2},
\end{eqnarray}
and
\begin{eqnarray}
Q = 2knN (N-1).
\end{eqnarray}
Again, this vanishes for a single M2-brane 
implying that our solution is physically meaningful only for $N \ge 2$.

\begin{figure}[t]
\begin{center}
\includegraphics[scale=.5]{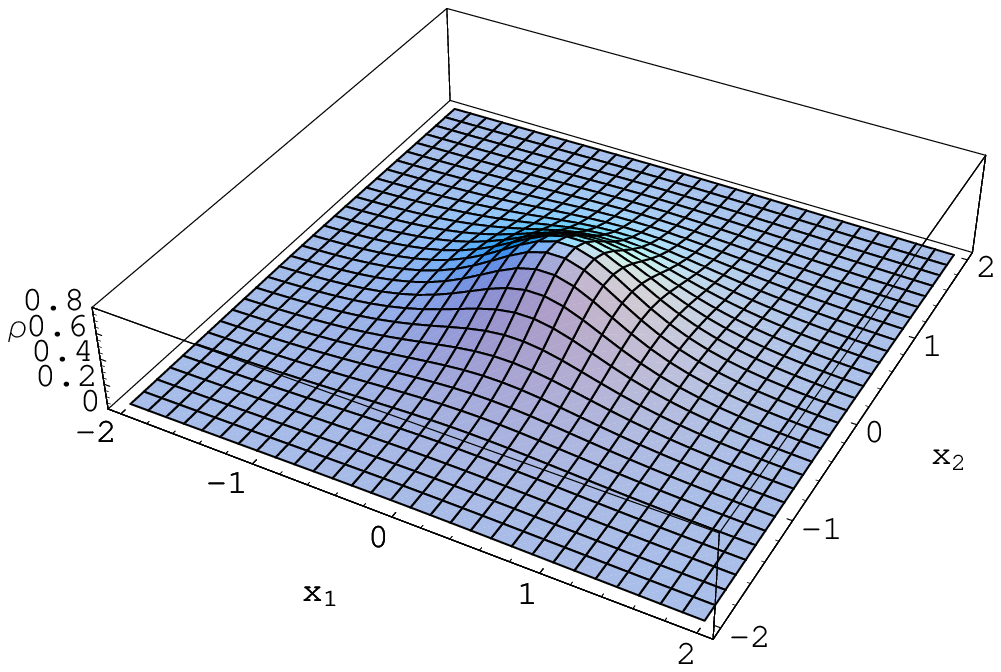}
\includegraphics[scale=.5]{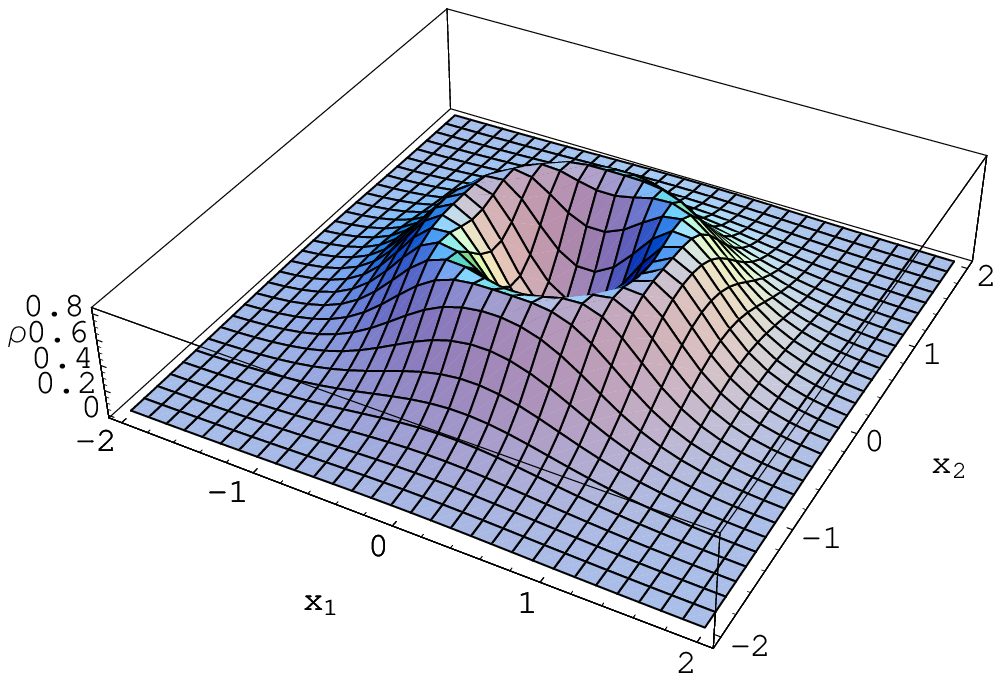}
\includegraphics[scale=.5]{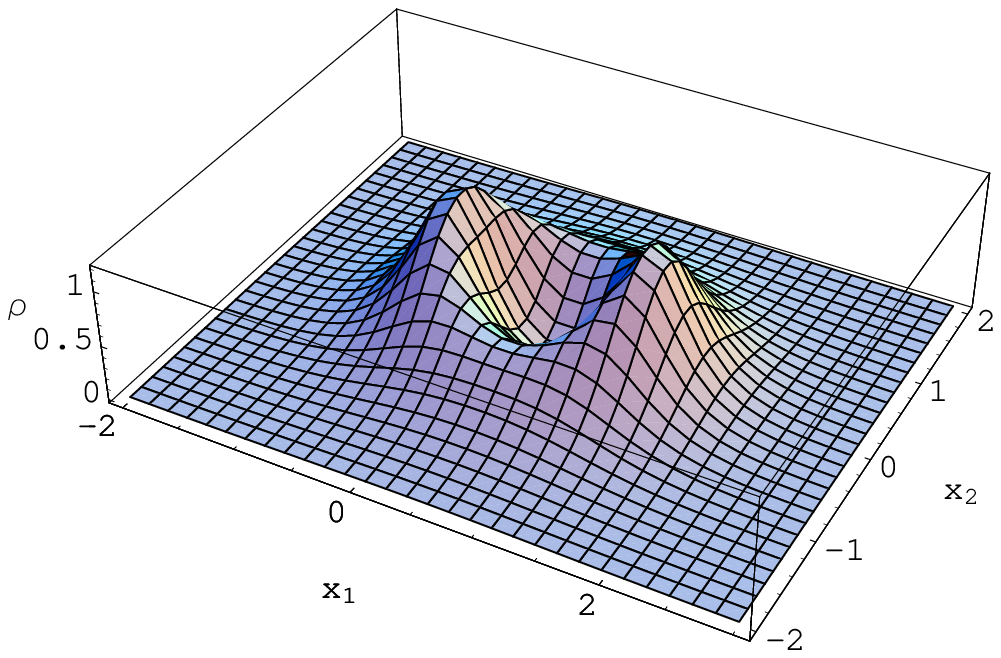}
\end{center}
\caption{
The profiles of the vortex solutions. 
$|\psi_z|^2$ is shown in the examples of (\ref{eqn:PowerProfile}) with $n=1$ (left) and 
$n=2$ (middle), and $f(z)=\frac{1}{z(z-1)}$ (right). }
\label{VortexProfiles}
\end{figure}

The phase $\theta$ is determined as follows.
For small and large values of $r$, $\rho$ behaves as
\begin{eqnarray}
& & \rho \sim r^{2 (n-1)}, \quad (r \to 0, n \ge 2), \\
& & \rho \sim r^{-2 n - 2}, \quad (r \to \infty),
\end{eqnarray}
and hence
\begin{eqnarray}
a_i (x) \sim - \partial_i \theta + (n-1) \epsilon_{ij} \frac{x^j}{r^2},
\quad (r \to 0).
\end{eqnarray}
The regularity of the gauge field at $r = 0$ demands 
$\theta = - (n-1) \arg z=-(n-1)\arctan(x_2/x_1)$. 
These are non-topological vortices since $|\psi_z|\rightarrow 0$ as 
$r\rightarrow \infty$.
To illustrate the solutions, profiles of $|\psi_z|^2$ are shown in Fig.\ref{VortexProfiles} 
for $f(z)=\frac 1z$, $\frac{1}{z^2}$ and $\frac{1}{z(z-1)}$, with $k=1$.

Instead of setting $w^{\dag\check A}=0$, we may set $z^{\hat A}=0$ and find similar solutions for
$w^{\dag\check A}$:
\bea
\psi_w(x)&=&e^{i\theta(x)}\rho^\half (x), 
\label{BPS_sol21}
\\
\rho (x) &=& -\frac{k}{2\pi} \nabla^2 \ln 
\left(
1 + |f(z)|^2
\right),
\label{BPS_sol22}
\\
\theta&=&(n-1)\arctan(x_2/x_1).
\label{BPS_sol23}
\eea
We call these solutions ``BPS-II."

A comment is in order regarding the relation between the solutions here and the ones found
in the relativistic ABJM model.
In \cite{Arai:2008kv}, the authors found 1/4 BPS vortex solutions in the 
F-term mass deformation of the relativistic ABJM model, where (similarly to our non-relativistic
case here) an abelian solution is embedded together with the fuzzy $S^3$ geometry.
Their analysis \cite{Arai:2008kv} relies on numerical study as there is no analytic solution 
known for relativistic Chern-Simons vortices, even for the abelian case. 
In contrast, in our non-relativistic case, the BPS equation reduces to the Liouville equation 
and is exactly solvable, as we have just shown. 
The solvability of the equation is a special feature of the non-relativistic limit of the 
Chern-Simons-matter theory.
The exact solutions (\ref{BPS_sol1}), (\ref{BPS_sol2}), ((\ref{BPS_sol21})-(\ref{BPS_sol23})) 
cannot be obtained from the relativistic ones. 

\section{The super Schr\"odinger symmetry preserved by the vortices}

The vortices found in the previous section are 
exact solutions to the BPS equations. 
In this section we study their supersymmetric properties and 
see how many of the non-relativistic supercharges are preserved by the 
BPS solutions.
Our notations and terminology of the non-relativistic SUSY transformations follow 
\cite{Lee:2009mm}.
We shall decompose the SUSY transformation parameters $\omega_{AB}$
and $\omega^{AB}$ using the basis $u_{\pm}$ in the same way as we did for the fermions:
\begin{eqnarray}
\omega &=& \tilde\omega_{-} u_{+} + \tilde\omega_{+} u_{-} 
= \frac{1}{\sqrt{2}} 
\left(
\begin{array}{c}
\tilde\omega_{-} + \tilde\omega_{+} \\
- i \tilde\omega_{-} + i \tilde\omega_{+}
\end{array}
\right), \\
\tilde\omega^{AB}_{\pm} &=& (\tilde\omega_{\pm AB})^{\dagger} 
= \frac{1}{2} \epsilon^{ABCD} \tilde\omega_{\pm CD},
\end{eqnarray}
where $\epsilon^{1234}=\epsilon_{1234}=1$.

The super Schr\"odinger symmetry is generated by 14 components of supercharges.
Ten of them are associated with {\it kinematical} SUSY, characterised by anti-commutation 
relations of the supercharges $\{Q_K, Q^\dag_K\}\sim{\C M}/m$ where ${\C M}$ is the total mass 
operator. 
The corresponding 10 SUSY parameters are
\begin{eqnarray}
\left(
\tilde\omega_{+ \hat{A} \hat{B}}, \;\;
\tilde\omega_{- \check{A} \check{B}},\;\;
\tilde\omega_{\pm \hat{A} \check{B}}
\right).
\end{eqnarray}
Two of the other supercharge components belong to {\it dynamical} SUSY, 
characterised by supercharge commutator $\{Q_D, Q^\dag_D\}\sim H$, and their SUSY parameters are
\begin{eqnarray}
\left(
\tilde\omega_{- \hat{A} \hat{B}}, \;\;
\tilde\omega_{+ \check{A} \check{B}}
\right).
\end{eqnarray}
The two remaining components are associated with {\it conformal} SUSY.

The transformation rules for the kinematical SUSY are
\begin{eqnarray}
\delta_K z^{\hat{A}} &=& \omega_{-}^{\hat{A} \hat{B}} \psi_{+ \hat{B}} - 
\omega_{+}^{\hat{A} \check{B}} \psi_{- \check{B}}, \label{eqn:KinSusy1}\\
\delta_K w^{\dagger \check{A}} &=& \omega_{-}^{\check{A} \hat{B}} \psi_{+ \hat{B}}
- \omega_{+}^{\check{A} \check{B}} \psi_{- \check{B}}, \\
\delta_K \psi_{+ \hat{A}} &=& - \omega_{+ \hat{A} \hat{B}} z^{\hat{B}} + 
\omega_{+ \hat{A} \check{B}} w^{\dagger \check{B}}, \\
\delta_K \psi_{- \check{A}} &=& \omega_{- \check{A} \hat{B}} z^{\hat{B}} 
+ \omega_{- \check{A} \check{B}} w^{\dagger \check{B}}, \\
\delta_K A_t &=& \frac{\pi \hbar}{km} 
\left[
z^{\hat{A}} \bar{\psi}_{+}^{\check{B}} \omega_{- \hat{A} \check{B}}
+ w^{\dagger \check{A}} \bar{\psi}_{+}^{\check{B}} \omega_{- \check{A} \check{B}}
+ z^{\hat{A}} \bar{\psi}_{-}^{\hat{B}} \omega_{+ \hat{A} \hat{B}} + 
w^{\dagger \check{A}} \bar{\psi}_{-}^{\hat{B}} \omega_{+ \check{A} \hat{B}}
\right. \nonumber \\
& & \qquad 
\left.
+ \omega_{-}^{\hat{A} \hat{B}} z^{\dagger}_{\hat{A}} \psi_{+ \hat{B}} 
+ \omega_{-}^{\check{A} \hat{B}} w_{\check{A}} \psi_{+ \hat{B}} 
+ \omega_{+}^{\hat{A} \check{B}} z^{\dagger}_{\hat{A}} \psi_{- \check{B}}
+ \omega_{+}^{\check{A} \check{B}} w_{\check{A}} \psi_{- \check{B}}
\right], \\
\delta_K A_{+} &=& \frac{2\pi}{km} 
\left(
w^{\dagger \check{A}} \bar{\psi}_{+}^{\check{B}} \omega_{+ \check{A} \check{B}}
+ \omega_{+}^{\hat{A} \hat{B}} \psi_{+ \hat{A}} z^{\dagger}_{\hat{B}}
\right), \\ 
\delta_K A_{-} &=& \frac{2\pi}{km} 
\left(
z^{\hat{A}} \bar{\psi}_{-}^{\hat{B}} \omega_{- \hat{A} \hat{B}} 
+ \omega_{-}^{\check{A} \check{B}} \psi_{- \check{A}} w_{\check{B}}
\right),
\end{eqnarray}
and the rules for the dynamical SUSY are 
\begin{eqnarray}
\delta_D z^{\hat{A}} &=& - \frac{i}{2m} \omega_{-}^{\hat{A} \hat{B}} 
D_{-} \psi_{+ \hat{B}}, \\
\delta_D w^{\dagger \check{A}} &=& \frac{i}{2m} \omega_{-}^{\check{A} 
\check{B}} D_{+} \psi_{- \check{B}}, \\
\delta_D \psi_{+ \hat{A}} &=& \frac{i}{2m} \omega_{- \hat{A} \hat{B}} 
D_{+} z^{\hat{B}}, \label{eqn:dynferm1}\\
\delta_D \psi_{- \check{A}} &=& - \frac{i}{2m} \omega_{+ \check{A} 
\check{B}} D_{-} w^{\dagger \check{B}}, \label{eqn:dynferm2}\\
\delta_D A_t &=& \frac{i \pi \hbar}{2 k m^2} 
\left[
- z^{\hat{A}} D_{+} \bar{\psi}_{-}^{\hat{B}} \omega_{- \hat{A} \hat{B}} 
- w^{\dagger \check{A}} D_{-} \bar{\psi}_{+}^{\check{B}} \omega_{+ \check{A} \check{B}} \right.\nn\\
&&\qquad\left.
+ \omega_{-}^{\hat{A} \hat{B}} w_{\check{A}} D_{+} \psi_{- \check{B}} 
+ \omega_{+}^{\hat{A} \hat{B}} z^{\dagger}_{\hat{A}} D_{-} \psi_{+ \hat{B}}
\right], \\
\delta_{D} A_{\pm} &=& 0.\label{eqn:DynSusy6}
\end{eqnarray}
For the sake of brevity we have used in these expressions rescaled SUSY parameters
\begin{eqnarray}
\left(
\omega_{+ \hat{A} \hat{B}}, \omega_{- \check{A} \check{B}}, 
\omega_{\pm \hat{A} \check{B}}
\right) 
&\equiv&
\sqrt{\frac{2mc}{\hbar}}
\left(
\tilde\omega_{+ \hat{A} \hat{B}}, \tilde\omega_{- \check{A} \check{B}}, 
\tilde\omega_{\pm \hat{A} \check{B}}
\right), \\
\left(
\omega_{+}^{\check{A} \check{B}}, \omega_{-}^{\hat{A} \hat{B}}, 
\omega_{\pm}^{\hat{A} \check{B}}
\right) 
&\equiv&
\sqrt{\frac{2mc}{\hbar}}
\left(
\tilde\omega_{+}^{\check{A} \check{B}}, \tilde\omega_{-}^{\hat{A} \hat{B}}, 
\tilde\omega_{\pm}^{\hat{A} \check{B}}
\right), 
\end{eqnarray}
and
\begin{eqnarray}
\left(
\omega_{- \hat{A} \hat{B}}, \omega_{+ \check{A} \check{B}}
\right)
&\equiv&
\sqrt{\frac{2m \hbar}{c}}
\left(
\tilde\omega_{- \hat{A} \hat{B}}, \tilde\omega_{+ \check{A} \check{B}}
\right), \\
\left(
\omega_{-}^{\check{A} \check{B}}, \omega_{+}^{\hat{A} \hat{B}}
\right)
&\equiv&
\sqrt{\frac{2m \hbar}{c}}
\left(
\tilde\omega_{-}^{\check{A} \check{B}}, \tilde\omega_{+}^{\hat{A} \hat{B}}
\right).
\end{eqnarray}
Dimensions of these new parameters are
\begin{eqnarray}
& & [\omega_{+ \hat{A} \hat{B}}] = [\omega_{- \check{A} \check{B}}] = 
[\omega_{\pm \hat{A} \check{B}}] = 
[\omega_{+}^{\check{A} \check{B}}] = [\omega_{-}^{\hat{A} \check{B}}] = 
[\omega_{\pm}^{\hat{A} \check{B}}] = 1, \\
& & [\omega_{- \hat{A} \hat{B}}] = [\omega_{+ \check{A} \check{B}}] =
[\omega_{-}^{\check{A} \check{B}}] = [\omega_{+}^{\hat{A} \hat{B}}] 
= M L.
\end{eqnarray}
We sketch derivation of the non-relativistic SUSY transformation formulae in the Appendix.

Let us first consider the BPS-I vortices, which are solutions to the BPS equations
\begin{eqnarray}
w^{\dagger \check{A}} = 0, \quad 
D_{-} z^{\hat{A}} = 0.
\label{eqn:BPS1cond}
\end{eqnarray}
Applying these conditions to the fermion transformation rules $\delta\psi$, we have
\begin{eqnarray}
\delta_K \psi_{+ \hat{A}} &=& - \omega_{+ \hat{A} \hat{B}} z^{\hat{B}}, \\
\delta_D \psi_{+ \hat{A}} &=& \frac{i}{2m} \omega_{- \hat{A} \hat{B}} D_{+} z^{\hat{B}}, \\
\delta_K \psi_{- \check{A}} &=& + \omega_{- \check{A} \hat{B}} z^{\hat{B}}, \\
\delta_D \psi_{- \check{A}} &=& 0,
\end{eqnarray}
hence the conditions $\delta \psi = 0$ imply
$\omega_{+ \hat{A} \hat{B}} = \omega_{- \hat{A} \hat{B}} = \omega_{- \check{A} \hat{B}} = 0$.
This means that the BPS-I solutions break 5 kinematical and 1 dynamical SUSYs.

For the BPS-II solutions the BPS equations are
\begin{eqnarray}
z^{\hat{A}} = 0, \quad D_{+} w^{\dagger \check{A}} = 0,
\label{eqn:BPS2cond}
\end{eqnarray}
and the transformation rules become
\begin{eqnarray}
\delta_K \psi_{+ \hat{A}} &=& \omega_{+ \hat{A} \check{B}} w^{\dagger \check{B}}, \\
\delta_D \psi_{+ \hat{A}} &=& 0, \\
\delta_K \psi_{- \check{A}} &=& \omega_{- \check{A} \check{B}} w^{\dagger \check{B}}, \\
\delta_D \psi_{- \check{A}} &=& - \frac{i}{2m} \omega_{+ \check{A} 
\check{B}} D_{-} w^{\dagger \check{B}}.
\end{eqnarray}
The conditions $\delta\psi=0$ then give 
$\omega_{+ \hat{A} \check{B}} = \omega_{- \check{A} \check{B}} 
= \omega_{+ \check{A} \check{B}} = 0$ and we see that the BPS-II solutions also break 5 kinematical and 1 dynamical SUSYs. 

The properties of the vortex solutions associated with the the conformal SUSY 
can be inferred from the fact that the conformal supercharge $S$ is written as a commutator 
of the special conformal generator $K$ and the dynamical supercharge $Q_D$
\cite{Leblanc:1992wu,Nakayama:2009cz,Lee:2009mm}, 
\beq
S=i[K,Q_D].
\eeq
Using the dynamical SUSY transformation rules (\ref{eqn:dynferm1}), (\ref{eqn:dynferm2})
we see that under the conformal SUSY
$\delta_S\psi_{+ \hat{A}}\sim \xi_{\hat A\hat B}z^{\hat B}$
and 
$\delta_S\psi_{- \check{A}}\sim \xi_{\check A\check B}w^{\dag\check B}$.
The former vanishes for the BPS-II conditions (\ref{eqn:BPS2cond}) 
whereas the latter vanishes for the BPS-I conditions (\ref{eqn:BPS1cond}). 
We may thus conclude that the BPS-I and BPS-II both preserve half of the conformal SUSY.
Note that once we turn on both $z^{\hat{A}}$ and $w^{\dagger \check{A}}$, 
the BPS equations break all the SUSYs in general and hence there would be 
only trivial solution $z^{\hat{A}} = w^{\dagger \check{A}} = 0$.
We summarize the results in table \ref{table:susy}.

\begin{table}[htbp]
\begin{center}
\begin{tabular}{c||c|c|c|c|c|c|c|c}
\hline
Type of & \multicolumn{4}{|c|}{Kinematical} & \multicolumn{2}{c|}{Dynamical}
& \multicolumn{2}{c}{Conformal}
\\
\cline{2-9}
SUSY & $\omega_{+ \hat{A} \check{B}}$ & $\omega_{+ \hat{A} \hat{B}}$ 
& $\omega_{- \check{A} \check{B}}$ & $\omega_{- \check{A} \hat{B}}$
& $\omega_{- \hat{A} \hat{B}}$ & $\omega_{+ \check{A} \check{B}}$ 
& $\xi_{\hat A\hat B}$ & $\xi_{\check A\check B}$
\\ 
\hline
BPS I & $\bigcirc$ & $\times$ & $\bigcirc$ & $\times$ & $\times$ & $\bigcirc$ &$\times$&$\bigcirc$
\\ 
\hline
BPS II & $\times$ & $\bigcirc$ & $\times$ & $\bigcirc$& $\bigcirc$& $\times$ &$\bigcirc$ &$\times$
\\
\hline
\end{tabular}
\end{center}
\caption{Broken and preserved SUSYs for our vortex solutions BPS-I and BPS-II. 
Here $\bigcirc$ for preserved, and $\times$ for broken SUSYs.}
\label{table:susy}
\end{table}
%

\section{Discussions}

In this paper we studied vortex solutions in the non-relativistic ABJM model and discussed the
non-relativistic SUSY they preserve. 
The ABJM model is a particularly interesting type of Chern-Simons-matter theory as its gravitational dual is well understood and its non-relativistic limit is also expected to have a gravitational dual through a non-relativistic version of AdS/CFT correspondence
\cite{Son:2008ye,Balasubramanian:2008dm,Maldacena:2008wh,Adams:2008wt,Herzog:2008wg}. 
We obtained exact solutions to the BPS equations and showed that these vortices preserve
half of the 10 kinematical, 2 dynamical and 2 conformal SUSYs. 
The solutions discussed in this paper are related to those of the Jackiw-Pi model.
In fact, the correspondence can be seen at the Lagrangian level.
Let us take the BPS-I ansatz for example:
setting $w_{\check A} = 0$ and assuming the fuzzy $S^3$ configuration,
\begin{eqnarray}
z^{\hat A} = \psi S^I, \quad A_{\mu} = a_{\mu} S^I S^{\dagger}_I, \quad 
\hat{A}_{\mu} = a_{\mu} S^{\dagger}_I S^I,
\end{eqnarray}
the non-relativistic ABJM model Lagrangian (\ref{eqn:NRABJMlag})
reduces to 
\beq
\mathcal{L}^{\rm NR, bos}_{\rm ABJM} = N (N-1) \mathcal{L}_{\rm JP},
\label{eqn:ABJM-JP}
\eeq
where
\beq
{\C L}_{\rm JP}=
\frac{k}{4\pi} \epsilon^{\mu \nu \lambda} a_{\mu} \partial_{\nu} 
a_{\lambda} 
+ \frac{i \hbar}{2} (- \psi \mathcal{D}_t \bar{\psi} + \bar{\psi} 
\mathcal{D}_t \psi) 
- \frac{\hbar^2}{2m} |\mathcal{D}_i \psi |^2 + \frac{\pi \hbar^2}{k 
m} (\psi \bar{\psi})^2,
\label{eqn:ABJM-JP2}
\eeq
is identified as the Lagrangian of the Jackiw-Pi model \cite{Jackiw:1990mb}.
We note that the fuzzy $S^3$ sphere ansatz is essential in this correspondence, and the 
correspondence holds only for $N\geq 2$.
The Jackiw-Pi model gives {\em abelian} vortices, whereas the gauge fields of the ABJM model 
(of $N\geq 2$) are non-abelian. 
We may say that the abelian vortices are embedded in the non-relativistic ABJM model,
with the non-abelian nature of the ABJM gauge fields converted into the fuzziness of the 
$S^3$ part and the numerical factor of (\ref{eqn:ABJM-JP}). 

While our solutions may be considered as an embedding of the abelian Jackiw-Pi vortices, 
it is not obvious from this fact alone how many of the non-relativistic 14 SUSYs are preserved by the BPS solutions. 
The Jackiw-Pi model $\mathcal{L}_{JP}$, which is the non-relativistic limit of 
the $\mathcal{N}=2$ abelian Chern-Simons-Higgs model \cite{Jackiw:1990pr}, 
does not exhibit 14 SUSYs but keeps only a part of them. 
This means that in order to see the full structure of the unbroken SUSY 
kept by the vortex solutions, it is necessary to analyse the BPS equation (78) derived from the 
original non-relativistic ABJM model, not the 
effective description (\ref{eqn:ABJM-JP}), (\ref{eqn:ABJM-JP2}).
One of 
our motivations to look for vortex solutions in the non-relativistic ABJM model arose from
their potential importance in holographic descriptions of $(1+2)$ dimensional condensed matter 
systems.
The structure of the preserved SUSYs is important for determining the corresponding 
solutions in the gravity side.
It would be interesting to find a solution that preserves 7 
Schr{\"o}dinger SUSYs in the eleven dimensional gravity dual.

Let us comment on more realistic models for 
condensed matter physics.
Physically interesting problems such as superconductivity and quantum Hall effect
involve external fields, and the parity of the systems is accordingly broken.
While the Jackiw-Pi vortex solutions that we described in this paper do not involve external
fields, a straightforward modification to include external fields is known once the
Lagrangian is suitably modified.
For example, let us add an additional term to the ABJM Lagrangian,
\beq
\delta{\C L}=\Tr \left[
F_{12} Z^{\hat A} Z^\dag_{\hat A} -\hat F_{12} Z^\dag_{\hat A} Z^{\hat A}\right].
\label{additional}
\eeq
With the fuzzy $S^3$ configuration 
\beq
F_{12}=B S^I S^\dag_I, \qquad
\hat F_{12}=B S^{\dag}_I S^I,
\eeq
together with the BPS-I ansatz, 
the Hamiltonian acquires an additional term proportional to 
$N(N-1)\frac{\hbar}{2m}B|\psi_z|^2$.
It is then possible to modify the vortex solutions to include the external fields following 
\cite{Ezawa:1991sh}.
It is interesting to see whether it is possible to accommodate more realistic models such as the 
Zhang-Hansson-Kivelson model \cite{Zhang:1988wy} of the quantum Hall effect.

Finally, it is also an interesting question whether the model allows other types of solitonic 
solutions, such as an embedding of non-abelian vortices, solutions with less supersymmetry, 
time-dependent solutions and so on.
For embedding non-abelian solutions, once one assumes an ansatz $A_{\mu} = \hat{A}_{\mu}$, 
the bi-fundamental scalar fields can be effectively treated as adjoint matter fields.
It would be interesting to see if it is possible to embed the
non-Abelian solutions of the Toda-type \cite{Dunne:1990qe}.
Finding more general solutions requires further study.
Determination of the complete moduli space of the solutions, in 
particular its relation to the broken SUSY structure, and clarification 
of the string theoretical origin of additional terms like (\ref{additional})
are also important problems.
We hope to come back to these issues in near future.

\subsection*{Acknowledgements}

We acknowledge helpful discussions with Claus Montonen, Seiji Terashima, 
Koji Hashimoto, Esko Keski-Vakkuri, Sean Nowling and Patta Yogendran. 
This work is in part supported by JSPS-Academy of Finland bilateral scientist exchange 
programme (SS) and the Academy of Finland Finnish-Japanese Core Programme Grant 
No. 112420 (SK).

\appendix

\section{The Non-relativistic supersymmetry}

In this appendix we describe how the non-relativistic SUSY transformation rules
(\ref{eqn:KinSusy1})-(\ref{eqn:DynSusy6}) arise in the non-relativistic limit of
the ${\C N}=6$ mass-deformed SUSY transformations. 
This is accomplished by decomposing the relativistic fields into non-relativistic particle 
and antiparticle parts, dropping the antiparticle part, and expand for large $c$ and $m$. 
Then the leading terms are identified as the kinematical and the next-to-leading as the
dynamical SUSY transformation terms.
See \cite{Nakayama:2009cz,Lee:2009mm,Leblanc:1992wu} for further details\footnote{
The literature available at the time of writing contains some mathematical typos.}, and
\cite{Duval:1994pw,Duval:1993hs} for related work on the Schr\"odinger and super Schr\"odinger
algebras.

We use the following conventions: 
the 3-dimensional gamma matrices are
\begin{eqnarray}
(\gamma^{\mu})_{\alpha} {}^{\beta} 
= (i \sigma_2, \sigma_1, \sigma_3), \quad 
\{\gamma^{\mu}, \gamma^{\nu} \} = 2 \eta^{\mu \nu}.
\end{eqnarray}
A spinor product is related to a matrix product as
\begin{eqnarray}
\Psi^{\dagger \alpha} \Psi_{\alpha} 
= -\mathbf{\Psi}^{\dagger} \gamma^0 \mathbf{\Psi},
\end{eqnarray}
where $\mathbf{\Psi}$ is a $2 \times 1$ matrix (vector) and the dagger in the right hand side is interpreted as the matrix adjoint.
In the following, we interpret $\Psi$ as the two component vector $\mathbf{\Psi}$.
The spinor indices are raised and lowered as 
\begin{eqnarray}
\theta^{\alpha} = \epsilon^{\alpha \beta} \theta_{\beta}, \quad 
\theta_{\alpha} = \epsilon_{\alpha \beta} \theta^{\beta}, \quad 
\epsilon^{12} = - \epsilon_{12} = 1.
\end{eqnarray}
The standard position of spinor contraction is 
\begin{eqnarray}
\theta \chi = \theta^{\alpha} \chi_{\alpha} 
= - \mathbf{\theta}^t \gamma^0 \mathbf{\chi}.
\end{eqnarray}

\subsection{The scalar part}

Let us start from the scalar part and consider the transformation
$\delta Y^A = i \omega^{AB} \Psi_B$. 
Using the fermion decomposition (\ref{eqn:FermDcp}) we may write
\begin{eqnarray}
(\omega^{AB})^{\alpha} (\Psi_B)_{\alpha} 
&=& - \omega^{AB} \gamma^0 \Psi_B \nonumber \\
&=& - \sqrt{\hbar c} i (\tilde\omega_{-}^{AB} \psi_{+ B}- \tilde\omega_{+}^{AB} \psi_{- B})
e^{- i \frac{mc^2}{\hbar} t}.
\end{eqnarray}
Decomposing the scalar field as
\begin{eqnarray}
Y^A = \frac{\hbar}{\sqrt{2m}} y^A
e^{- \frac{mc^2}{\hbar} t} 
+ \textrm{(anti particle)},
\end{eqnarray}
and dropping the antiparticle part, 
the first two components of the SUSY transformation becomes 
\begin{eqnarray}
\delta z^{\hat{A}} &=& 
\frac{\sqrt{2m \hbar c}}{\hbar}
\left(
\tilde\omega_{-}^{\hat{A} \hat{B}} \psi_{+ \hat{B}}
+ \tilde\omega_{-}^{\hat{A} \check{B}} \psi_{+ \check{B}}
- \tilde\omega_{+}^{\hat{A} \hat{B}} \psi_{- \hat{B}} 
- \tilde\omega_{+}^{\hat{A} \check{B}} \psi_{- \check{B}}
\right) 
\nonumber \\
&=& 
\frac{\sqrt{2m \hbar c}}{\hbar}
\left(
\tilde\omega_{-}^{\hat{A} \hat{B}} \psi_{+ \hat{B}}
+ \frac{i \hbar}{2mc} \tilde\omega_{-}^{\hat{A} \check{B}} D_{+} \psi_{- \check{B}}
- \frac{i \hbar}{2mc} \tilde\omega_{+}^{\hat{A} \hat{B}} D_{-} \psi_{+ \hat{B}} 
- \tilde\omega_{+}^{\hat{A} \check{B}} \psi_{- \check{B}}
\right) \nn\\
&&\qquad + \textrm{ (higher order terms)}.
\end{eqnarray}
We have used the Dirac equations (\ref{fermioneq1}), (\ref{fermioneq2}) to go to the second line.
From the leading order we find (using the rescaled parameters),
\beq
\delta_K z^{\hat{A}} = \omega_{-}^{\hat{A} \hat{B}} \psi_{+ \hat{B}}
- \omega_{+}^{\hat{A} \check{B}} \psi_{- \check{B}}, 
\eeq
and from the next-to-leading order,
\beq
\delta_D z^{\hat{A}} = - \frac{i}{2m} \omega_{-}^{\hat{A} \hat{B}} 
D_{-} \psi_{+ \hat{B}}.
\eeq
From the other components we similarly find
\begin{eqnarray}
\delta_K w^{\dagger \check{A}} &=& 
\omega_{-}^{\check{A} \hat{B}} \psi_{+ \hat{B}}
- \omega_{+}^{\check{A} \check{B}} \psi_{- \check{B}}, \\
\delta_D w^{\dagger \check{A}} &=& \frac{i}{2m} \omega_{-}^{\check{A} \check{B}}
D_{+} \psi_{- \check{B}}.
\end{eqnarray}

\subsection{The fermion part}

Next we consider the transformation of the fermion. 
The first term on the RHS 
of the SUSY transformation 
can be written upon particle-antiparticle decomposition (and
neglecting the antiparticle) as
\begin{eqnarray}
&&\gamma^{\mu} \omega_{AB} D_{\mu} Y^B \nn\\
&&\quad= 
- i \frac{mc}{\sqrt{2m}} \gamma^0 \omega_{AB} y^B e^{- \frac{mc^2}{\hbar} t} 
+ \frac{1}{\sqrt{2m}} \frac{\hbar}{c} \gamma^0 \omega_{AB} D_t y^B 
e^{- \frac{mc^2}{\hbar} t}
+ \frac{\hbar}{\sqrt{2m}} \gamma^i \omega_{AB} D_i y^B 
e^{- \frac{mc^2}{\hbar} t}\nn\\
&&\quad= 
\left(
\begin{array}{c}
\frac{mc}{\hbar} (\tilde\omega_{+AB} - \tilde\omega_{-AB}) y^B 
+ \frac{1}{c}(i\tilde\omega_{+AB} - i\tilde\omega_{-AB}) D_t y^B 
+ i\tilde\omega_{+AB} D_{-} y^B - i\tilde\omega_{-AB} D_{+} y^B
\\
\frac{mc}{\hbar} (i\tilde\omega_{+AB} + i\tilde\omega_{-AB}) y^B 
- \frac{1}{c}(\tilde\omega_{+AB} + \tilde\omega_{-AB}) D_t y^B 
+ \tilde\omega_{+AB} D_{-} y^B + \tilde\omega_{-AB} D_{+} y^B
\end{array}
\right) \nn\\
&&\qquad\qquad\qquad\qquad \times \frac{\hbar}{2\sqrt{m}} 
e^{- i \frac{mc^2}{\hbar} t}.
\end{eqnarray}
The $D_t y^B $ terms are subleading and can be dropped. 
The mass independent part in the 
second term on the RHS is also subleading. 
The mass-dependent term gives non-trivial contributions, 
\beq
\frac{mc}{\hbar}Y^{C}\omega_{\hat A C}
=\frac c2 \sqrt m\left(
\begin{array}{cc}
z^{\hat B} \tilde\omega_{-\hat A\hat B}+w^{\dag\check B}\tilde\omega_{-\hat A\check B}
+z^{\hat B}\tilde\omega_{+\hat A\hat B}+w^{\dag\check B}\tilde\omega_{+\hat A\check B}\\
-iz^{\hat B} \tilde\omega_{-\hat A\hat B}-iw^{\dag\check B}\tilde\omega_{-\hat A\check B}
+iz^{\hat B}\tilde\omega_{+\hat A\hat B}+iw^{\dag\check B}\tilde\omega_{+\hat A\check B}
\end{array}
\right).
\eeq
As the fermion transformations decompose as
\begin{eqnarray}
\delta \Psi_A 
= \frac{1}{\sqrt{2}} \sqrt{\hbar c} 
\left(
\begin{array}{c}
\delta \psi_{-A} + \delta \psi_{+A} \\
- i \delta \psi_{-A} + i \delta \psi_{+A}
\end{array}
\right)
e^{- i \frac{mc^2}{\hbar} t},
\end{eqnarray}
we find
\begin{eqnarray}
\delta \psi_{-\hat A} + \delta \psi_{+\hat A}  &=& + i \frac{mc}{\sqrt{2m \hbar c}} 
\left( - i \tilde\omega_{-\hat{A} \hat{B}} + i \tilde\omega_{+ \hat{A} \hat{B}} \right)
z^{\hat{B}} 
+ i \frac{mc}{\sqrt{2 m \hbar c}} 
\left( - i \tilde\omega_{- \hat{A} \check{B}} + i \tilde\omega_{+ \hat{A} \check{B}} 
\right) 
w^{\dagger \check{B}} \nonumber \\
& & - \frac{\hbar}{\sqrt{2 m \hbar c}} 
\left(
- i \tilde\omega_{-\hat{A} \hat{B}} D_{+} z^{\hat{B}} - i \tilde\omega_{- \hat{A} 
\check{B}} D_{+} w^{\dagger \check{B}} + i \tilde\omega_{+ \hat{A} \hat{B}} 
D_{-} z^{\hat{B}} + i \tilde\omega_{+ \hat{A} \check{B}} D_{-} w^{\dagger \check{B}}
\right) \nonumber \\
& & - \frac{mc}{\hbar} \frac{\hbar}{\sqrt{2 m \hbar c}} 
\left(
z^{\hat{B}} \tilde\omega_{- \hat{A} \hat{B}} + w^{\dagger \check{B}} \tilde\omega_{- 
\hat{A} \check{B}} + z^{\hat{B}} \tilde\omega_{+ \hat{A} \hat{B}} + w^{\dagger 
\check{B}} \tilde\omega_{+ \hat{A} \check{B}}
\right), \\
- i \delta \psi_{-\hat A} + i \delta \psi_{+\hat A} &=& + i \frac{mc}{\sqrt{2m \hbar c}} 
\left( - \tilde\omega_{-\hat{A} \hat{B}} - \tilde\omega_{+ \hat{A} \hat{B}} \right)
z^{\hat{B}} 
+ i \frac{mc}{\sqrt{2 m \hbar c}} 
\left( - \tilde\omega_{- \hat{A} \check{B}} - \tilde\omega_{+ \hat{A} \check{B}} 
\right) 
w^{\dagger \check{B}} \nonumber \\
& & - \frac{\hbar}{\sqrt{2 m \hbar c}} 
\left(
\tilde\omega_{-\hat{A} \hat{B}} D_{+} z^{\hat{B}} + \tilde\omega_{- \hat{A} 
\check{B}} D_{+} w^{\dagger \check{B}} +  \tilde\omega_{+ \hat{A} \hat{B}} 
D_{-} z^{\hat{B}} +  \tilde\omega_{+ \hat{A} \check{B}} D_{-} w^{\dagger \check{B}}
\right) \nonumber \\
& & - \frac{mc}{\hbar} \frac{\hbar}{\sqrt{2 m \hbar c}} 
\left(
- i z^{\hat{B}} \tilde\omega_{- \hat{A} \hat{B}} - i w^{\dagger \check{B}} \tilde\omega_{- 
\hat{A} \check{B}} + i z^{\hat{B}} \tilde\omega_{+ \hat{A} \hat{B}} + i w^{\dagger 
\check{B}} \tilde\omega_{+ \hat{A} \check{B}}
\right).
\end{eqnarray}
Because of the Dirac equations (\ref{fermioneq1}), (\ref{fermioneq2}), 
$\delta \psi_{- \hat{A}}$ on the LHS is subleading.
Then in terms of the rescaled SUSY parameters we obtain the kinematical 
$\delta_K$ and dynamical $\delta_D$ SUSY transformations
\begin{eqnarray}
\delta_K \psi_{+\hat{A}} &=& - \omega_{+ \hat{A} \hat{B}} z^{\hat{B}} 
+ \omega_{+ \hat{A} \check{B}} w^{\dagger \check{B}}, \\
\delta_D \psi_{+ \hat{A}} &=&  \frac{i}{2m} \omega_{- \hat{A} \hat{B}} 
D_{+} z^{\hat{B}}.
\end{eqnarray}

Similarly, 
\begin{eqnarray}
\delta \psi_{- \check{A}} + \delta \psi_{+ \check{A}}
&=& \frac{i \hbar}{\sqrt{2 m \hbar c}} 
\left(
\tilde\omega_{- \check{A} \hat{B}} D_{+} z^{\hat{B}} 
+ \tilde\omega_{- \check{A} \check{B}} D_{+} w^{\dagger \check{B}}
- \tilde\omega_{+ \check{A} \hat{B}} D_{-} z^{\hat{B}} - \tilde\omega_{+ \check{A} 
\check{B}} D_{-} w^{\dagger \check{B}}
\right)
\nonumber \\
& & +\frac{2mc}{\sqrt{2m \hbar c}} 
\left(
\tilde\omega_{- \check{A} \hat{B}} z^{\hat{B}} 
+ \tilde\omega_{- \check{A} \check{B}} w^{\dagger \check{B}}
\right), \\
- i \delta \psi_{- \check{A}} + i \delta \psi_{+ \check{A}}
&=&- \frac{ \hbar}{\sqrt{2 m \hbar c}} 
\left(
\tilde\omega_{ \check{A} \hat{B}} D_{+} z^{\hat{B}} 
+ \tilde\omega_{- \check{A} \check{B}} D_{+} w^{\dagger \check{B}}
+ \tilde\omega_{+ \check{A} \hat{B}} D_{-} z^{\hat{B}} + \tilde\omega_{+ \check{A} 
\check{B}} D_{-} w^{\dagger \check{B}}
\right)
\nonumber \\
& & - i \frac{2mc}{\sqrt{2m \hbar c}} 
\left(
\tilde\omega_{- \check{A} \hat{B}} z^{\hat{B}} 
+ \tilde\omega_{- \check{A} \check{B}} w^{\dagger \check{B}}
\right),
\end{eqnarray}
and again due to the Dirac equations (\ref{fermioneq1}), (\ref{fermioneq2}) we may
drop $\delta \psi_{+ \check{A}}$ on the LHS, leading to
\begin{eqnarray}
\delta_K \psi_{- \check{A}} &=& \omega_{- \check{A} \hat{B}} z^{\hat{B}}
+ \omega_{- \check{A} \check{B}} w^{\dagger \check{B}}, \\
\delta_D \psi_{- \check{A}} &=& - \frac{i}{2m} \omega_{+ \check{A} 
\check{B}} D_{-} w^{\dagger \check{B}}.
\end{eqnarray}

\subsection{The gauge field part}

Finally we consider the gauge field part.
Note that the temporal and the spatial parts of the relativistic SUSY 
transformation formula come with different powers of $c$:
\begin{eqnarray}
\delta A_t &=& + \frac{2\pi}{k \hbar} 
\left(
Y^A \Psi^{\dagger B} \gamma^0 \omega_{AB} 
+ \omega^{AB} \gamma^0 \Psi_A Y^{\dagger}_B
\right), \\
\delta A_{\pm} &=& - \frac{2\pi}{k \hbar c} 
\left(
Y^A \Psi^{\dagger B} \gamma^{\pm} \omega_{AB} 
+ \omega^{AB} \gamma^i \Psi_A Y^{\dagger}_B
\right), 
\end{eqnarray}
where  
\begin{eqnarray}
\gamma^{\pm} \equiv \gamma^1 \pm i \gamma^2 =
\left(
\begin{array}{cc}
\pm i & 1 \\
1 & \mp i
\end{array}
\right).
\end{eqnarray}
Upon non-relativistic decomposition of the fields the temporal part becomes
\bea
\delta A_t
&=& \frac{2\pi}{k} \sqrt{\frac{\hbar c}{2m}}
\left[
- \frac{i \hbar}{2mc} z^{\hat{A}} D_{+} \bar{\psi}_{-}^{\hat{B}} 
\tilde\omega_{- \hat{A} \hat{B}} 
- \frac{i \hbar}{2mc} w^{\dagger \check{A}} D_{+} \bar{\psi}_{-}^{\hat{B}} 
\tilde\omega_{- \check{A} \hat{B}}
+ z^{\hat{A}} \bar{\psi}_{+}^{\check{B}} \tilde\omega_{- \hat{A} \check{B}} 
+ w^{\dagger \check{A}} \bar{\psi}_{+}^{\check{B}} \tilde\omega_{- \check{A} \check{B}}
\right. \nonumber \\
& &\qquad 
+ z^{\hat{A}} \bar{\psi}_{-}^{\hat{B}} \psi_{+ \hat{A} \hat{B}} 
+ w^{\dagger \check{A}} \bar{\psi}_{-}^{\hat{B}} \tilde\omega_{+ \check{A} \hat{B}}
- \frac{i \hbar}{2mc} z^{\hat{A}} D_{-} \bar{\psi}_{+}^{\check{B}} 
\tilde\omega_{+ \hat{A} \check{B}} 
- \frac{i \hbar}{2mc} w^{\dagger \check{A}} D_{-} 
\bar{\psi}_{+}^{\check{B}} \tilde\omega_{+ \check{A} \check{B}}
\nonumber \\
& &\qquad
+ \tilde\omega_{-}^{\hat{A} \hat{B}} z^{\dagger}_{\hat{A}} \psi_{+ \hat{B}}
+ \tilde\omega_{-}^{\check{A} \hat{B}} w_{\check{A}} \psi_{+ \hat{B}}
+ \frac{i \hbar}{2mc} \tilde\omega_{-}^{\hat{A} \check{B}} z^{\dagger}_{\hat{A}}
D_{+} \psi_{- \check{B}} + \frac{i \hbar}{2mc} \tilde\omega_{-}^{\check{A} 
\check{B}} w_{\check{A}} D_{+} \psi_{- \check{B}}
\nonumber \\
& &\qquad \left.
+ \frac{i \hbar}{2mc} \tilde\omega_{+}^{\hat{A} \hat{B}} z^{\dagger}_{\hat{A}} 
D_{-} \psi_{+ \hat{B}} + \frac{i \hbar}{2mc} \tilde\omega_{+}^{\check{A} 
\hat{B}} w_{\check{A}} D_{-} \psi_{+ \hat{B}}
+ \tilde\omega_{+}^{\hat{A} \check{B}} z^{\dagger}_{\hat{A}} \psi_{- \check{B}}
+ \tilde\omega_{+}^{\check{A} \check{B}} w_{\check{A}} \psi_{- \check{B}}
\right] \nonumber \\
& &\qquad + \textrm{(higher order terms)}.
\eea
Using the rescaled SUSY parameters we obtain
\begin{eqnarray}
\delta_K A_t &=& \frac{\pi \hbar}{km} 
\left[
z^{\hat{A}} \bar{\psi}_{+}^{\check{B}} \omega_{- \hat{A} \check{B}}
+ w^{\dagger \check{A}} \bar{\psi}_{+}^{\check{B}} \omega_{- \check{A} \check{B}}
+ z^{\hat{A}} \bar{\psi}_{-}^{\hat{B}} \omega_{+ \hat{A} \hat{B}} + 
w^{\dagger \check{A}} \bar{\psi}_{-}^{\hat{B}} \omega_{+ \check{A} \hat{B}}
\right. \nonumber \\
& & \qquad 
\left.
+ \omega_{-}^{\hat{A} \hat{B}} z^{\dagger}_{\hat{A}} \psi_{+ \hat{B}} 
+ \omega_{-}^{\check{A} \hat{B}} w_{\check{A}} \psi_{+ \hat{B}} 
+ \omega_{+}^{\hat{A} \check{B}} z^{\dagger}_{\hat{A}} \psi_{- \check{B}}
+ \omega_{+}^{\check{A} \check{B}} w_{\check{A}} \psi_{- \check{B}}
\right], \\
\delta_D A_t &=& \frac{i \pi \hbar}{2 k m^2} 
\left[
- z^{\hat{A}} D_{+} \bar{\psi}_{-}^{\hat{B}} 
\omega_{- \hat{A} \hat{B}} - w^{\dagger \check{A}} 
D_{-} \bar{\psi}_{+}^{\check{B}} \omega_{+ \check{A} \check{B}} 
+ \omega_{-}^{\hat{A} \hat{B}} w_{\check{A}} D_{+} 
\psi_{- \check{B}} + \omega_{+}^{\hat{A} \hat{B}} 
z^{\dagger}_{\hat{A}} D_{-} \psi_{+ \hat{B}}
\right].
\nonumber \\
\end{eqnarray}

The spatial part of the transformation formula can be found similarly. 
From
\begin{eqnarray}
\delta A_{+} &=& \frac{4\pi}{kc} \sqrt{\frac{\hbar c}{2m}} 
\left[
- \frac{i \hbar}{2mc} z^{\hat{A}} D_{+} \bar{\psi}_{-}^{\hat{B}} 
\tilde\omega_{+ \hat{A} \hat{B}} 
- \frac{i \hbar}{2mc} w^{\dagger \check{A}} D_{+} \bar{\psi}_{-}^{\hat{B}}
\tilde\omega_{+ \check{A} \hat{B}}
+ z^{\hat{A}} \bar{\psi}_{+}^{\check{B}} \tilde\omega_{+ \hat{A} \check{B}}
+ w^{\dagger \check{A}} \bar{\psi}_{+}^{\check{B}} \tilde\omega_{+ 
\check{A} \check{B}}  
\right. \nonumber \\
& & \qquad \qquad \left.
+ \tilde\omega_{+}^{\hat{A} \hat{B}} \psi_{+ \hat{A}} z^{\dagger}_{\hat{B}} 
+ \tilde\omega_{+}^{\hat{A} \check{B}} \psi_{+ \hat{A}} w_{\check{B}} 
+ \frac{i \hbar}{2mc} \tilde\omega_{+}^{\check{A} \hat{B}} D_{+} \psi_{- 
\check{A}} z^{\dagger}_{\hat{B}} 
+ \frac{i \hbar}{2mc} \tilde\omega_{+}^{\check{A} \check{B}} D_{+} \psi_{- 
\check{A}} w_{\check{B}}
\right] \nonumber \\
& & \qquad + \textrm{(higher order terms)},
\end{eqnarray}
we obtain
\begin{eqnarray}
\delta_K A_{+} &=& \frac{2\pi}{km} 
\left(
w^{\dagger \check{A}} \bar{\psi}_{+}^{\check{B}} \omega_{+ \check{A} \check{B}}
+ \omega_{+}^{\hat{A} \hat{B}} \psi_{+ \hat{A}} z^{\dagger}_{\hat{B}}
\right), \\
\delta_D A_{+} &=& 0, 
\end{eqnarray}
and from
\begin{eqnarray}
\delta A_{-} &=& 
\frac{4\pi}{kc} \sqrt{\frac{\hbar c}{2m}} 
\left[
 z^{\hat{A}} \bar{\psi}_{-}^{\hat{B}} \tilde\omega_{- \hat{A} \hat{B}}
+ w^{\dagger \check{A}} \bar{\psi}_{-}^{\hat{B}} \tilde\omega_{- \check{A} 
\hat{B}} 
- \frac{i \hbar}{2mc} z^{\hat{A}} D_{-} \bar{\psi}_{+}^{\check{B}} 
\tilde\omega_{- \hat{A} \check{B}} - \frac{i \hbar}{2mc} w^{\dagger \check{A}} 
D_{-} \bar{\psi}_{+}^{\check{B}} \tilde\omega_{- \check{A} \check{B}}
\right. \nonumber \\
& & \qquad \qquad \left.
+ \frac{i \hbar}{2mc} \tilde\omega_{-}^{\hat{A} \hat{B}} D_{-} \psi_{+ \hat{A}} z^{\dagger}_{\hat{B}}
+ \frac{i \hbar}{2mc} \tilde\omega_{-}^{\hat{A} \check{B}} D_{-} \psi_{+ 
\hat{A}} w_{\check{B}} 
+ \tilde\omega_{-}^{\check{A} \hat{B}} \psi_{- \check{A}} z^{\dagger}_{\hat{B}}
+ \tilde\omega_{-}^{\check{A} \check{B}} \psi_{- \check{A}} w_{\check{B}}
\right] \nonumber \\
& & \qquad + \textrm{(higher order terms)},
\end{eqnarray}
we have
\begin{eqnarray}
\delta_K A_{-} &=& \frac{2\pi}{km} 
\left(
z^{\hat{A}} \bar{\psi}_{-}^{\hat{B}} \omega_{- \hat{A} \hat{B}} 
+ \omega_{-}^{\check{A} \check{B}} \psi_{- \check{A}} w_{\check{B}}
\right), \\
\delta_D A_{-} &=& 0.
\end{eqnarray}

\providecommand{\href}[2]{#2}\begingroup\raggedright\endgroup

\end{document}